\documentclass[preprint]{aastex63} 

\usepackage{graphicx}
\usepackage{amsmath,amssymb}

\shorttitle{Contact binaries observed spectroscopically}
\shortauthors{Rucinski}

\begin{document}

\title{Lessons from the high-resolution spectroscopy of AW~UMa and $\epsilon\,$CrA: \\
Is the Lucy model valid?}

\author{Slavek M. Rucinski} 
\affil{Department of Astronomy and Astrophysics, 
University of Toronto \\
50 St. George St., Toronto, Ontario, M5S~3H4, Canada}
\email{slavek.rucinski@utoronto.ca}

\begin{abstract}
A re-examination of high-resolution spectral monitoring of
the W~UMa-type binaries AW~UMa and $\epsilon\,$CrA casts doubt
on the widely utilized \citet{Lucy1968a,Lucy1968b} model of contact binaries. 
The detection of the very faint profile of the secondary component in AW~UMa 
leads to a new spectroscopic determination of the mass ratio, 
$q_{\rm sp} =0.092 \pm 0.007$, which is close to the previous,
medium-resolution spectroscopic result of \citet{PR2008}, 
$q_{\rm sp}= 0.101 \pm 0.006$, and remains  substantially different from 
a cluster of generally accepted photometric results by several
authors, concentrated around $q_{\rm ph}=0.080 \pm 0.005$.
A survey of binaries with the best-determined values of the mass ratio 
shows a common tendency for  $q_{\rm ph} < q_{\rm sp}$. 
The tendency for systematically smaller values of 
$q_{\rm ph}$ may result from the overfilling of the primary
lobe and under-filling of the secondary lobe relative to the Roche model
geometry, as predicted by the \citet{Step2009} model.  
Despite the observed moderate inter-systemic velocities, the photometric
Lucy model may remain useful in providing approximate, though 
biased, results for the mass ratio. A complicating factor
in detailed spectral analysis may be the occurrence of Enhanced Spectral-line
Perturbations (ESP) projected over the secondary profiles, appearing in different 
numbers in the two studied binaries. The ESPs are tentatively identified within 
the St\c{e}pie\'{n} model as collision fronts or fountains of  primary-component 
gas from the circumbinary, energy-carrying flow.
\end{abstract}
    
 
%\bigskip

\vfil

\section{Introduction}     % ============================================   Sec.1
\label{intro}

This paper is a combined re-discussion of the high spectral-resolution
time monitoring of two W~UMa-type binary stars,  
AW~UMa \citep[paper P1]{Rci2015} and 
$\epsilon\,$CrA \citep[paper P2]{Rci2020}. 
The binaries are located at the high-mass, long-period end of the 
W~UMa-binary effective-temperature sequence.
Excellent recent introductions to the current state of  research on W~UMa-type or
contact binaries can be found in \citet{GS2008} and \citet{Gaz2024}, while
a large collection of data and derived, Lucy-model dependent quantities 
is in \citet{Lat2021}.

The W~UMa binaries,  sometimes also called ``contact binaries'',
consist of apparently Main Sequence stars in the spectral range 
from early F-type to early K-type. 
W~UMa binaries are moderately common in the solar
vicinity, with one such system per about 500 F-K dwarfs \citep{Rci2002a}. 
Their kinematic and metallicity properties are characteristic 
of the old disk population \citep{Rci2013a}.       %   Rci2013a - metallicity, Rci2013b - MOST
The high-luminosity detection advantage of  the early
F-type systems, such as AW~UMa and $\epsilon\,$CrA,
compensates for their smaller numbers compared with somewhat later
counterparts and for a tendency to have more dissimilar components, resulting in  
small variability amplitudes.  
The mass ratios $q = M_2/M_1$ tend to peak 
in a wide range around $q \simeq 0.3 - 0.5$ for more common late-F and early-G type
binaries, but rarer early-F systems tend to show small values of $q$ and consequently 
small light variations. 
There may exist a bifurcation in the mass ratio as a function of the
orbital period \citep[Fig.6]{Rci2010}; this matter is currently
a subject of further research \citep{Gaz2024}.
% One well-documented case of a merger of a rather massive system
%  into a single star has been observed \citep{Tyl2011}. 

The similarity of our two targets was not accidental but resulted 
from a limited availability of telescopes equipped with efficient high-resolution 
spectrographs that could be used for extended time-monitoring programs.
The two objects in question are the brightest W~UMa binaries in the sky 
(with magnitudes $V = 6.8$ and 4.8) making them suitable for observation 
with a signal-to-noise per spectral-resolution element of around  
one hundred over a few consecutive nights. Such time-monitoring is 
resource-expensive, and it is possible that the two objects 
will remain the only ones observed in this manner 
for the foreseeable future. This paper brings a comparison and rediscussion of
the results from P1 and P2, highlighting that rapid spectral monitoring 
is one of the few avenues for advancing 
the currently somewhat stalled research on the validity of the 
popular \citet{Lucy1968a,Lucy1968b} model for W~UMa-type binaries 
\citep{Webb2003}, especially considering the new, attractive, yet unstudied 
model proposed by \citet{Step2009}; see also \citet{Step2013}.
 
In both of the considered here binaries,
%AW~UMa and $\epsilon\,$CrA, 
the early-F type, more massive 
primary component\footnote{It is proposed to avoid the
photometric definition of the ``primary component'' 
in favor of the more physical one based on the mass.}
dominates over its smaller companion in terms of the mass and size.  
Because of its small mass, approximately ten times smaller in both cases,
the secondary is an energetically inert appendage to the primary. It
carries the binary angular momentum, produces strong tidal effects 
and provides a substantial surface area for the combined radiative losses of
the system. 
The two binaries are typical for the W~UMa-type binaries in that 
their secondary components appear as hot as the primaries. 
This characteristic is unique among binary stars with unequal mass 
components and is the main defining feature of the W~UMa-type binaries.

The small masses and  unexpectedly high surface brightnesses of the secondary 
components  in apparently stable W~UMa-type systems 
emphasize the need to determine the mass ratios $q$ with high accuracy. 
Ideally, these determinations
should involve as few initial assumptions as possible. Currently, many
determinations of the mass ratios are based on light-curve-synthesis 
photometric solutions which utilize the relatively complex 
\citet{Lucy1968a,Lucy1968b} model, where $q_{\rm ph}$ is one of its many parameters. 
In contrast, spectroscopic determinations, through 
a ratio of orbital semi-amplitudes, $q_{\rm sp} = K_1/K_2$, 
are expected to reflect global binary dynamics directly
but are observationally more challenging to determine.

This paper focuses particularly on the observational properties of
the secondary components and the energy-transport mechanisms that lead
to their high surface brightnesses. We also  
address the  good agreement between the photometric and 
spectroscopic determinations for $\epsilon\,$CrA at $q = 0.13$ (Section~\ref{sec-eps}) 
and the continuing substantial discrepancy for AW~UMa (Section~\ref{sec-aw}).
For AW~UMa, the medium-resolution result of  \citet{PR2008},  
$q_{\rm sp}= 0.101 \pm 0.006$, significantly differs from the 
widely accepted $q_{\rm ph} \simeq 0.08$, based on several photometric
studies starting from \citet{MD1972a} and ending with \citet{Wilson2008}.

We note that before P1, there have been several previous short-duration attempts to 
observe AW~UMa spectroscopically
\citep{BeP1964,McL1981,And1983,Rens1985,Rci1992}, These efforts 
mainly focused on {\it detecting\/}  the spectroscopically
faint secondary component and confirming the photometric mass ratio.
The DDO analysis of \citet{PR2008} was the first to provide a larger amount of
data. In contrast, $\epsilon\,$CrA was analyzed spectroscopically using a 
cross-correlation function (CCF) 
approach only once by \citet{GD1993}, followed by the high-quality
high-resolution spectral analysis reported in P2. 
The previous line-by-line measurements 
for the primary component of $\epsilon\,$CrA by \citet{TW1975} 
currently have a very limited value.

A general description and comparison of both binaries 
%AW~UMa and $\epsilon\,$CrA 
in terms of their general properties is in Section~\ref{both}.
Sections~\ref{sec-eps} and \ref{sec-aw} discuss the complex appearance
of the secondary components as seen in spectroscopy. In these sections,
the new term Enhanced Spectral-line Perturbation (ESP) is introduced to describe 
radial-velocity-defined features of increased density of the 
primary-component gas, projected onto the secondary component profile. 
A newly detected, very faint signature of the secondary component in 
AW~UMa confirms the ``large'' value of $q_{\rm sp}$ of about 0.10. 

Section~\ref{discr} demonstrates that the tendency 
for $q_{\rm ph} < q_{\rm sp}$ is a common feature in the best 
mass-ratio determinations of other W~UMa binaries. 
The previously unrecognized weaknesses of the Lucy model 
in providing unbiased values of $q_{\rm ph}$ are discussed in 
Section~\ref{mod-q}. As discussed in Section~\ref{model}, 
the tendency for $q_{\rm ph} < q_{\rm sp}$
cannot be explained by the Lucy model but is consistent
with the \citet{Step2009} model. Section~\ref{concl} discusses the
implications of the presented results and suggests possible directions 
for future research. 

% Latex spelling:   St\c{e}pie\'{n} model 
  
%\vfil 

\bigskip

\section{The two binaries as seen in radial velocities}   % ======================   Sec.2
\label{both}

All results presented in papers P1 and P2 were based on 
the radial velocity (RV) profiles obtained through the analysis of individual spectra  
using the linear, Broadening-Functions (BF) deconvolution technique
\citep{Rci2002b,Rci2010,Rci2012}. The technique is computationally 
more demanding than the inherently non-linear,
easier-to-implement Cross-Correlation Function 
(CCF) technique; it  provides high-quality RV profiles in reference
to the shape and strength of the spectral lines for a star of the same spectral type.
The CCF technique is simpler but adequate for measurements of well-peaked
mean profiles; line-by-line RV measurements are practically impossible for
W~UMa-type binaries due to extreme broadening and blending of the spectral lines.
The BF profiles should not be interpreted as RV-projected images of a binary system;
to do so, an assumption of a solid-body rotation is necessary; generally, we do not
know the spatial location of a given RV feature. 
Strict solid-body rotation is a basic assumption of the Lucy model, 
yet determining the correct model is precisely the main subject of our investigation.

As with other spectral deconvolution process, the BF technique has the limitation
of being sensitive to the presence of dispersed matter in the binary system: 
Any {\it  emission\/} component in observed line spectra would produce 
a dip or would weaken the restored velocity profile, while any {\it absorption\/} 
would produce stronger features in BF profiles. In general,  
deviations from a standard, stellar-atmosphere optical-depth dependence
of the source function can modify the profile. The
BF technique, thanks to its linear properties,  tends to show such
deviations more directly than the CCF technique.

%  Figure:  -----   BF Eps CrA at phases 0.25 and 0.75 over-plotted ---------------    Fig.1
\begin{figure}[t]
\begin{center}
\includegraphics[width=11.5cm]{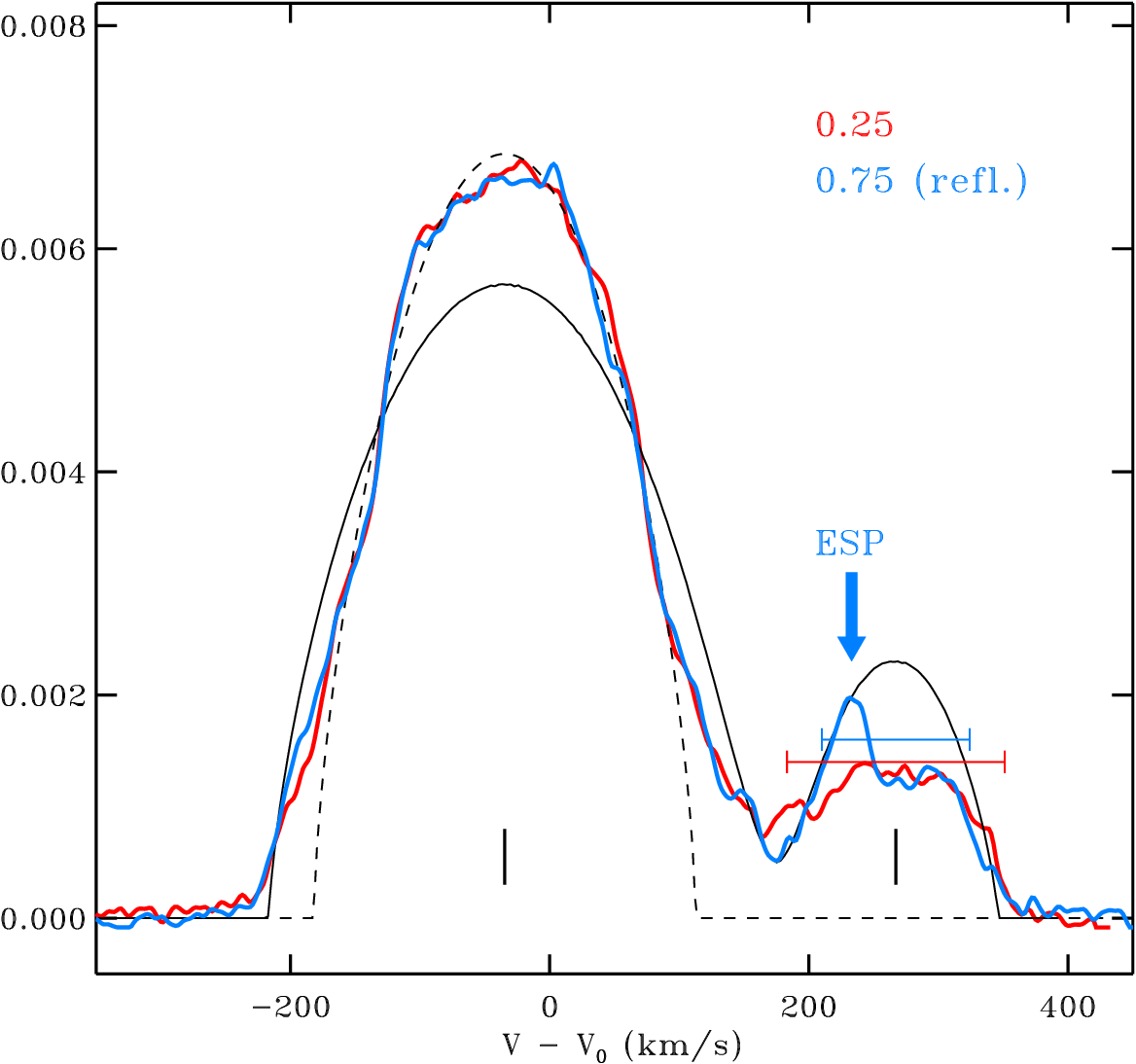}          % Eps CrA 0.25/0.75 over-plotted
\caption{
\footnotesize
The BF profiles  
for $\epsilon\,$CrA at both orbital quadratures, i.e.\ phases $\phi = 0.25$ (red) and
$\phi = 0.75$ (blue), in units per a RV sampling interval of 8.5 km~s$^{-1}$. 
The profile $\phi = 0.75$ is inverted in the velocity scale %for over-plotting 
with the common origin at the RV of the binary mass center at
$V_0 = -62.54$ km~s$^{-1}$. 
The Lucy-model profile for 
%an F2V spectral type at 
the orbital phase $\phi = 0.25$ and the assumed mass ratio $q = 0.13$, 
the fill-out parameter $f = 0.25$ is shown by a thin black line. 
Vertical dashes mark the mass-center RV's while   
a thin, broken line shows the rotational profile fit to the primary component. 
The red and blue bars over the 
secondary component show the widths of the secondary profile observed 
at both orbital quadratures. An Enhanced Spectral-line Perturbation (ESP)
affecting the secondary profile at $\phi = 0.75$ 
is pointed by an arrow at 235 km~s$^{-1}$. 
}
\label{fig_BFquad}
\end{center}
\end{figure}
% ------------------------------------------------------------------------------------------

The data for AW~UMa (P1) and $\epsilon\,$CrA  (P2) were analyzed 
with the spectra resampled to the same spectral resolution of 8.5 km~s$^{-1}$ 
(the resolving power $R \simeq 35,\!000$), with 
the final RV profiles for $\epsilon\,$CrA  being of much better quality: The median 
signal-to-noise ratio per resolution element 
of $S/N=57$ was for AW~UMa (P1) and $S/N=120$ for $\epsilon\,$CrA (P2). 
We show two typical RV profiles for $\epsilon\,$CrA at both orbital quadratures 
% ($\phi = 0.25$ and $\phi = 0.75$)  
in Figure~\ref{fig_BFquad}. Typical profiles for AW~UMa are very similar, but
are much more complex and variable within the secondary component parts
and exhibit larger observational noise.

%\break        % a new page
%\vfill
\bigskip

\noindent
Here is a summary of similarities and differences between the two binaries:
\begin{enumerate}
% 1.
\item In both binaries, the upper two-thirds of the primary component profile is narrower than 
predicted by the Lucy contact-binary model but agrees very well
with a standard profile for a single star of a smaller radius 
 rotating at the orbit-synchronous rotation rate (sizes 88\% and
85\% of the critical Roche-lobe side radius for AW~UMa and $\epsilon\,$CrA,
respectively).
This property was recognized in AW~UMa already by \citet{Rens1985}
using photographic spectra.  
The base of the primary profile in both binaries, referred to as
the ``pedestal'' in P1, is roughly as wide, as predicted by the Lucy model. 
The narrow primary profile may be due to the anticyclonic circulation
around the rotation pole of the mass-losing primary, 
as predicted by the hydrodynamic model  of \citet{Oka2002}, 
while the wide pedestal is explained by the 
binary circulation model of \citet{Step2009}. These observational properties
contribute to our discussion of an appropriate model for
W~UMa binaries in Section~\ref{model}. 
% 2.
\item In both binaries, the primary-component profile widths show a small
tidal distortion (line-of-centers elongation)  
of the more massive star of $<\!5$~km~s$^{-1}$.
This number should be compared with the mean values of 
$V \sin i \simeq 181$~km~s$^{-1}$
for AW~UMa (P1, Fig.~3) and  $V \sin i \simeq 147$~km~s$^{-1}$ for $\epsilon\,$CrA
(P2, Fig.~3). These width variations are partly masked by greater noise in AW~UMa
but are better defined for $\epsilon\,$CrA. 
% 3.
\item The primary component of AW~UMa showed a network of ``ripples'' or
non-radial pulsations partly extending into the profile pedestal. 
The ripples, which are visible after subtraction of the stable 
mean profile (Figures~8, 9, 10 in P1), look very similar to the non-radial 
pulsations detected (subsequently to the work reported in P1) 
in the rapidly rotating, bright star Altair 
($\alpha$~Aql, A7V, $V \sin i \simeq 240$ km~s$^{-1}$  
\citep{Rieu2023,Rieu2024}) which is a $\delta$~Sct-type star.
So far, there has been no attempt to detect photometrically these small, 
rapid pulsations in AW~UMa.
The primary of $\epsilon\,$CrA appears to be entirely free of 
non-radial pulsations; they would be easily detectable because of the higher 
$S/N$ of its spectra.
% 4.
\item      % slightly rewritten
The secondary-component RV profiles of both binaries show 
significant deviations from the predictions of the Lucy model. 
They appear much weaker and flatter, which may indicate 
either a genuine faintness of the secondaries or a different (later) 
spectrum that does not match the template used in the BF determination.
% 5.        % the old #5 split into two, so that #6 and the following have new numbers
\item The secondary-component profiles in both binaries were distorted by 
narrow perturbations of increased intensity of absorption lines. 
The disturbances, referred to as ``wisps'' in P1 and P2 due to their visibility 
in time sequences of the RV profiles, systematically
 shifted with the orbital phase, complicating our 
interpretation of the secondary component profile in AW~UMa.
See Figure~13 in P1 for an illustration of their confusing appearance. 
The wisps do not appear to result from the Struve-Sahade effect 
 \citep{Gies1997,Linder2007}, which is observed in early-type stars. The S-S effect 
describes the increased strength of spectral lines of the approaching star
over half of the orbital period.  In contrast,
the wisps in AW~UMa and $\epsilon\,$CrA 
are narrow, well-localized in radial velocities, and 
their phase migratation  allows their approximate geometric 
 localization within the binary structure (Sections~\ref{sec-eps} \& \ref{sec-aw}).
% new 6.
 \item          % previously 2nd half of #5, added explanations
$\epsilon\,$CrA (P2) showed one well defined, strong, narrow ``wisp''. However, 
an identical, strong feature appeared also in AW~UMa -- 
albeit among typically three to four wisps present there at a time.  
The phase evolution of this prominent 
wisp can be interpreted as due to a localized region of 
the  primary-component gas overlying the secondary component
profile. Its location in both stars was estimated at the sub-observer 
orbital phase $\phi = 0.65$ and may correspond to a collision front of the gas 
striking the primary after circulating the secondary component. 
The disturbance was very stable and remained 
visible in $\epsilon\,$CrA during the whole observing run of the 
28 orbital revolutions of the binary.  
It is marked by an arrow in the $\phi = 0.75$ profile in Figure~\ref{fig_BFquad};
its orbital-phase evolution is discussed in Section~\ref{sec-eps}. 
In this paper, we introduce the term  
Enhanced Spectral-line Perturbations (ESP) to stress that such 
perturbations  in Broadening Function profiles correspond to stronger
absorption spectra due to spacially-localized gas atoms having the same 
atomic excitation properties as those of the early-F type primary star. 
% new 7.
\item The better data quality and somewhat simpler profiles
permitted the detection of weakly defined outermost 
edges of the secondary-component velocity profile
in $\epsilon\,$CrA (P2).  Since both edges were detected,  
the velocity field on the secondary seemed to be well confined, 
similar to that of a detached star. The mean velocity from the edges -- 
assumed to be that of the secondary mass center --
showed an approximately sine-curve phase 
dependence, as expected in anti-phase to that observed for the primary component. 
The semi-amplitude $K_2$ derived in this way 
%  = 267.1$ km~s$^{-1}$,
% together with the very well defined $K_1 =  $ of the primary, 
led to a mass ratio  
$q_{\rm sp} = K_1/K_2 = 0.130 \pm 0.001$, in general consistency with the 
photometric Lucy-model estimates, $q_{\rm ph} = 0.114 \pm 0.003$ \citep{Tw1979}
and $0.1244 \pm 0.0014$ \citep{WR2011}.  
Unfortunately, a similar direct determination of $K_2$ for AW~UMa in P1 
could not be done due to the simultaneous presence of several 
ESPs affecting the secondary-component profile.
Indirect estimates in P1, based on the poorly-defined 
mean profile shape suggested a value of the mass ratio 
$q_{\rm sp}  \simeq 0.10$.  
% new 8.
\item The edge-to-edge widths of the  $\epsilon\,$CrA secondary-component RV profile 
appeared to differ for the two orbital quadratures. 
Expressed as the observed half-widths, they were approximately 
90 km~s$^{-1}$ at $\phi = 0.25$ and 70 km~s$^{-1}$ 
$\phi = 0.75$. 
This unexplained variability is marked by horizontal bars over the 
secondary-component profiles in Figure~\ref{fig_BFquad}. This subject is discussed
further in Section~\ref{sec-eps}.   
% new 9.
\item No parallel photometry was conducted during the spectroscopic programs described
in P1 and P2. Information restored from the available spectra
(Figure~2 in P1; Figure~13 in P2) reveals different and complex phase variations of
the integrated strength of all spectral lines in the used BF window
for the two binaries. %AW~UMa and $\epsilon\,$CrA.
% new 10.
\item The spectroscopic data do not provide information on the
orbital inclinations of either binary. 
A difference in the photometric variability amplitudes suggests that the
$\epsilon\,$CrA orbit is viewed at a slightly more inclined angle than AW~UMa.
The tentative assumptions of the previous, Lucy-model photometric estimates
(Sec.~\ref{intro}) are 
$i \simeq 78^\circ - 80^\circ$ for AW~UMa and $i \simeq 72^\circ - 73^\circ$ for
$\epsilon\,$CrA, though these are not explicitly used in the paper. It may be useful
to note that for $\epsilon\,$CrA, we can possibly observe the volume between 
the components during the transit eclipse ($\phi = 0$), 
with a line-of-sight passing ``above'' the secondary component 
(see Figure~\ref{img_EpsCrA_Lucy}).
% new 11.
\item Although they have the same spectral type and a similar, small
mass ratio, the binaries are definitely {\it not identical\/}: 
Their orbital periods differ substantially: 0.4387 days for AW~UMa and 
0.5914 days for $\epsilon\,$CrA. The 
observed maximum-light $V$ magnitudes are 6.84 and 4.74, and the 
parallax data, 14.78 mas and 31.81 mas for AW~UMa and $\epsilon\,$CrA
respectively \citep{Gaia2020}. These values 
give the absolute maximum-light magnitudes $M_V = 2.69$ for AW~UMa
and 2.25 for $\epsilon\,$CrA. 
The longer period  and the higher luminosity of $\epsilon\,$CrA align 
with its larger primary mass, as discussed in P1 and P2.
% new 12.
\item Both binaries appear to systematically change their orbital periods, showing 
well-defined parabolic variations in the $O-C$ eclipse moments, but {\it in
the opposite sense\/}: In AW~UMa the period shortens 
\citep{Rci2013b},                     % MOST result
while in $\epsilon\,$CrA the period lengthens (P2). The respective 
timescales are $(-2.3 \pm 0.1) \times 10^6$ yrs 
              and $(+1.26 \pm 0.04) \times 10^6$ yrs. 
AW~UMa has a physical, G7V proper-motion companion $67\arcsec$ 
away ($>\!4,600$ AU), 
while no third body has been detected for $\epsilon\,$CrA so far.
              
\end{enumerate}

% ===========  Section: Secondary of Eps CrA ==========================   Sec.3

\bigskip
%\vfill

\section{The secondary component of  $\epsilon\,$CrA}
\label{sec-eps}

The secondary components keep the secret of the moderate mass transfer 
and large energy transfer between the component stars
in W~UMa-type systems such as $\epsilon\,$CrA 
and AW~UMa. These processes are critical to understanding  
the relatively long duration of the contact phase in these systems. 
The following sections will describe the observed properties 
of the secondary components, beginning with $\epsilon\,$CrA and then moving 
on to AW~UMa. This sequence deviates from the analysis order in the 
previous studies (P1 and P2), prioritizing the simpler and more thoroughly 
observed $\epsilon\,$CrA to more effectively elucidate the shared features. 
For $\epsilon\,$CrA, we rely on the previously established orbital semi-amplitudes, 
$K_1$ and $K_2$, as reported  in P2, without any modifications. 
The focus will be on the lessons learned from the case of $\epsilon\,$CrA and
how they could be applied to understanding AW~UMa, as discussed in Section~\ref{sec-aw}.

As described in P2, the secondary component 
was relatively easy to detect in the radial velocity profiles of $\epsilon\,$CrA. 
While no distinct peak or profile centroid could be detected, 
the range or extent of the profile was well defined by ``edges'' 
on images derived from the RV profiles. 
The presence of such RV edges made the secondary component 
profiles dramatically different from that predicted by the Lucy model and
resembling those for a velocity field on a detached binary component. 

The  profiles were significantly fainter and flatter than those expected by the 
Lucy model (Figure~\ref{fig_BFquad}). However, the extremes of the
observed profiles were consistently well-defined to assume that their mean values
could correspond to the motion of the secondary mass center. These values
varied in anti-phase to the orbital motion of the primary component, allowing 
the derivation of the semi-amplitude $K_2$. This finding stands in stark 
contrast to the complex appearance of the secondary-component profiles
in AW~UMa (P1, Figure~13). However, the differences in data quality and 
quantity between the studies of $\epsilon\,$CrA and AW~UMa could influence 
this comparison, potentially favoring $\epsilon\,$CrA due to its better 
data quality and phase coverage.

%  -------   Figure:   RV ranges &  drift of the RV feature in Eps CrA  -----------------------------   Fig.2
\begin{figure}[ht]
\begin{center}
\includegraphics[width=12.5cm]{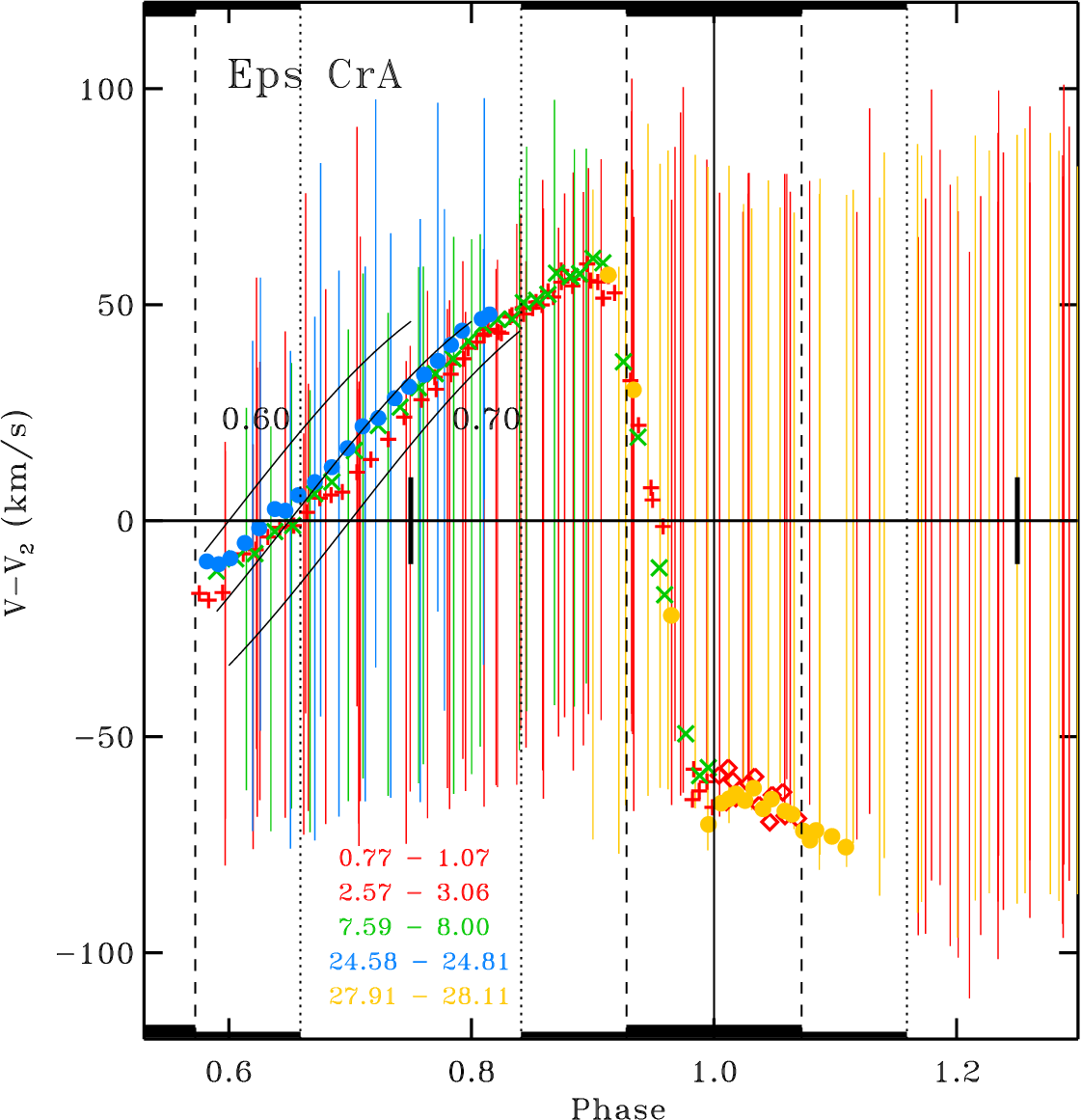}            %  Eps CrA sec RV ranges
\caption{
\footnotesize
The thin vertical bars show measurements of
the edge-to-edge extent of the secondary component RV profiles 
for $\epsilon\,$CrA, expressed relative to the velocity of the secondary component's 
mass center. These repeated and difficult  estimates
were derived from images formed from individual Broadening Function  profiles. 
%such as Figure~\ref{fig_Eps_jump}.
The broken wavy lines give the expected range of velocities
for the inner critical Roche lobe of the mass ratio $q = 0.130$.
Observations from individual nights are color-coded as per the
legend in the figure, where the integer part of the phase gives 
the number of elapsed orbits from the assumed initial epoch (see P2).
Thick black lines at the upper and lower figure frames indicate 
the phase ranges of the estimated total and partial eclipses. 
In the left part of the figure, the phase drift of the dominant 
Enhanced Spectral-line Perturbation (ESP) is represented by colored symbols 
according to the figure legend. 
The inclined  lines alongside these symbols represent the predicted 
phase drifts for three sub-observer longitudes, expressed in orbital phases: 
0.6, 0.65, and 0.7.  
}
\label{fig_Eps_spot}
\end{center}
\end{figure}
% -------------------------------------------------------------------------------------------

The  two main complications in the secondary-component profiles of $\epsilon\,$CrA 
were: 
(1)~The half-extends or half-widths of the profile 
(functionally similar to $V \sin i$ in rotating stars)
showed variations with phase, and (2)~a moving, phase-dependent, 
``spiky'' feature,  what we call ``Enhanced Spectral-line Perturbation'' (ESP) 
was observed in the orbital phases $0.5 < \phi <  1.0$ (see Figure~\ref{fig_BFquad}).  
The first complication                                % of the varying width of the profile, 
was noted only briefly in P2.
It is shown in Figure~\ref{fig_Eps_spot} as
the full measured extent of the secondary component profile, relative
to the assumed sine-curve motion of the secondary with $K_2 = 267$ km~s$^{-1}$.
The widths were determined from images similar to Figures 5 -- 9 in P2, rather than
from individual RV profiles; this provided a better definition of the measurements. 
The phase range in Figure~\ref{fig_Eps_spot} 
is limited to the phases when the secondary component 
is expected to be in front of the primary, $0.58 < \phi < 1.42$.  
Trends in the secondary-component RV widths in 
Figure~\ref{fig_Eps_spot} appear to be well established:
The profile is noticeably narrower ($\simeq  70$ km~s$^{-1}$)
around $\phi \simeq 0.75$ and wider ($\simeq  90$ km~s$^{-1}$)
around $\phi \simeq 0.25$ where the velocity range is similar to that
for a solid-body rotation of the critical Roche-model lobe.
The deviations from a sine-curve, while relatively small (20 -- 30 km~s$^{-1}$)
compared with the large value of $K_2$, are definitely systematic. Their presence does
not affect the essential anti-symmetric phase variation relative to the primary, yet 
it is concerning. 
We do not have a simple explanation for them beyond expectations that the velocity
field over the secondary component may be complex and that
the outer layers of the secondary are seen at constantly changing orientation and thus 
oblique-view, optical-depth accumulation.

% ------------------  Image:  appearance of  Eps CrA  - Lucy model -----------------  Fig.3
\begin{figure}[h]
\begin{center}
\includegraphics[width=12.5cm]{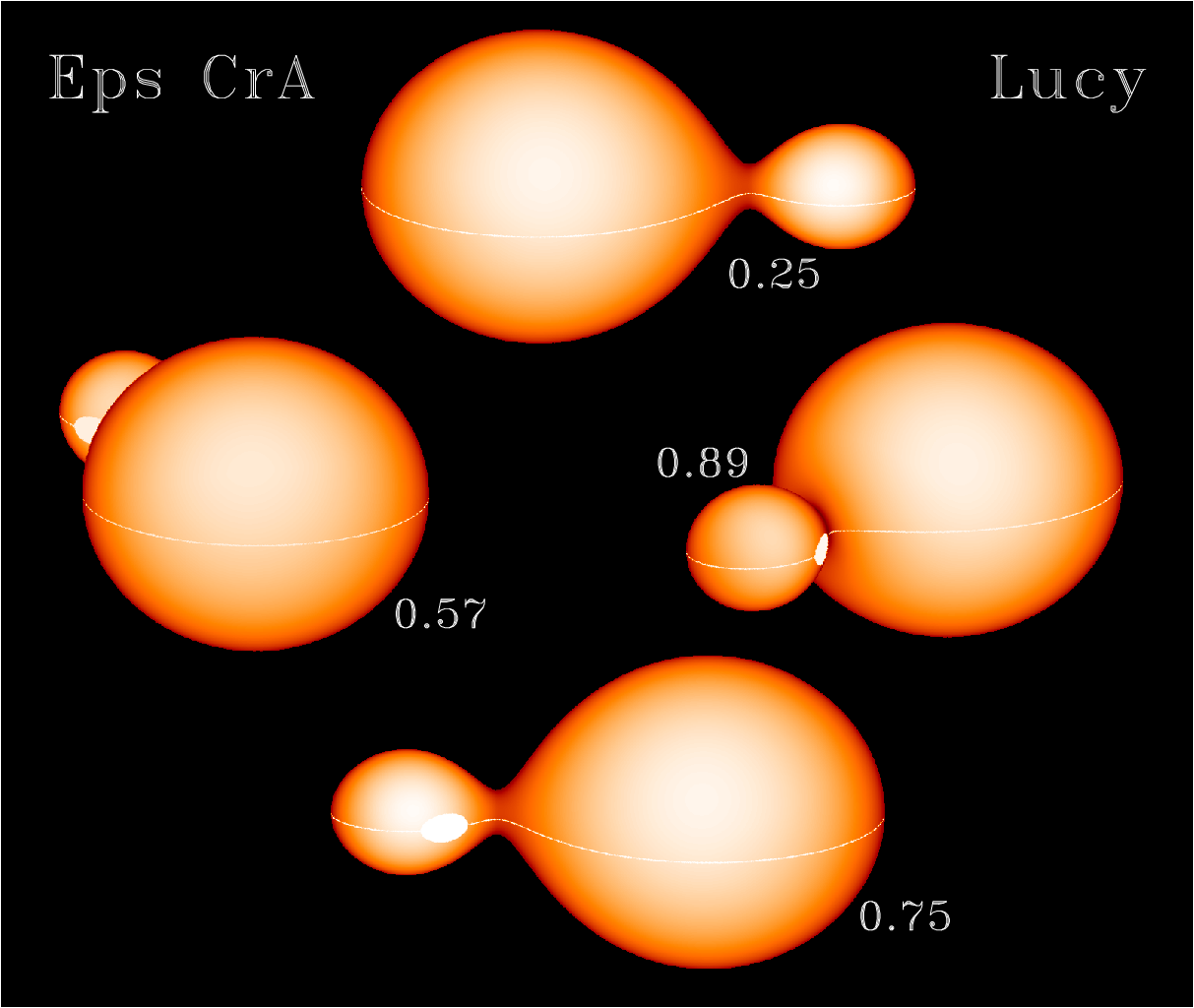}        % images of Eps CrA - Lucy model
\caption{
\footnotesize
A  schematic view of $\epsilon\,$CrA at four orbital phases, as indicated in the figure.
The Lucy model with the mass ratio $q=0.13$, the fill-out 
parameter $f=0.25$ and the orbital inclination $i=73$ degrees were assumed in
generating these images. The 
orbital phases $\phi \simeq 0.57$ and $\phi \simeq 0.89$ correspond to the orientations
when the region of the Enhanced Spectral-line Perturbation (ESP),  
shown as a white patch, comes into view or is about to
disappear behind the edge of the secondary component. Except for the
approximate location in longitude, the geometry and extent of the ESP-causing region 
in the stellar latitude are unknown and arbitrary in the figure (see the text).
}
\label{img_EpsCrA_Lucy}
\end{center}
\end{figure}
% ----------------------------------------------------------------------------------------

The second complication of the $\epsilon\,$CrA secondary-component profile,
the prominent ESP visible in the phase range $0.57 < \phi < 0.89$ (Figure~\ref{fig_Eps_spot}) 
requires special attention due to its location, strength and persistence. 
During each observed binary orbit, the region emerged at $\phi \simeq 0.57$ 
with the RV around $-10$ km~s$^{-1}$ relative to the secondary center,
then drifted to positive velocities reaching about +55 km~s$^{-1}$ at  
$\phi \simeq 0.89$ where it started overlapping with the primary component profile. 
Constraints on the ESP location within the binary can be derived utilizing the orbital-phase 
dependence of the RV variations. 
The simplest model would be a region anchored to the secondary 
component, with the observed phase variations caused by 
the variable component of the observer-directed product $\vec{r} \,\Omega$. 
Here, $\vec{r}$ gives the spatial position of the active
ESP region relative to the secondary center and $\Omega$ is 
the angular rotational velocity of the binary.  
The inclined lines in Figure~\ref{fig_Eps_spot} give predictions of such a phase
dependence for an ESP located on the equator of the secondary component
of the inner critical Roche equipotential 
%(passing through the $C_1$ point) 
at three sub-observer points, $\phi = 0.60, 0.65, 0.70$.
The zero RV crossing by the drift line 
%(Figure~\ref{fig_Eps_spot}) 
suggests a position on the secondary component at the sub-observer position 
$\phi \simeq 0.65$. 
Lucy-model generated images of $\epsilon\,$CrA with the ESP located at that longitude are 
shown for four orbital phases in Figure~\ref{img_EpsCrA_Lucy}.

To be detected as a positive intensity deviation in the BF, the gas producing 
the feature must possess properties similar to those of the outer 
layers of the primary component, 
with the same atomic-excitation properties. This is a direct result of the
BF determination process which used a high-temperature, early-F type spectrum
as a reference.
The ESP is visible on the hemisphere centered at $\phi = 0.75$, which 
is on the opposite side to where a Coriolis-force-deflected flow
from the primary would be expected. Therefore, the stream of matter 
must have circled the secondary component to collide 
with some obstacle, forming a fountain or a collision front. We return to
this subject in the discussion of the contact binary model in Section~\ref{model}. 

The ESP strength did not seem to vary with the orbital phase within the 
observed interval $0.57 < \phi < 0.89$ 
and remained at an intensity of about 1.5\% to 2\% ($\pm 0.2\%$) 
relative to the mean integrated Broadening 
Function\footnote{The BF integral was close to unity 
through the selection of the stellar template of a spectral type similar
to that of the primary component; see P2 and the lower panel of Fig.~13 there.}. 
It was also surprisingly stable over the entire duration of our
$\epsilon\,$CrA  campaign, remaining at the same place on five observing 
nights within 28 binary orbits, i.e.\ for over 2 weeks. 
Although the active region was placed on the equipotential 
surface in Figure~\ref{img_EpsCrA_Lucy}, its spatial location is not well
defined; it may be in the space overlying the secondary, 
possibly as in Figure~\ref{fig_Stepien} in the discussion of the St\c{e}pie\'{n} model  
in Section~\ref{mod-kst}.
 
 %  -------   Figure:       the feature jump during transit in Eps CrA  ------------  Fig.4
\begin{figure}[h]
\begin{center}
\includegraphics[width=13.5cm]{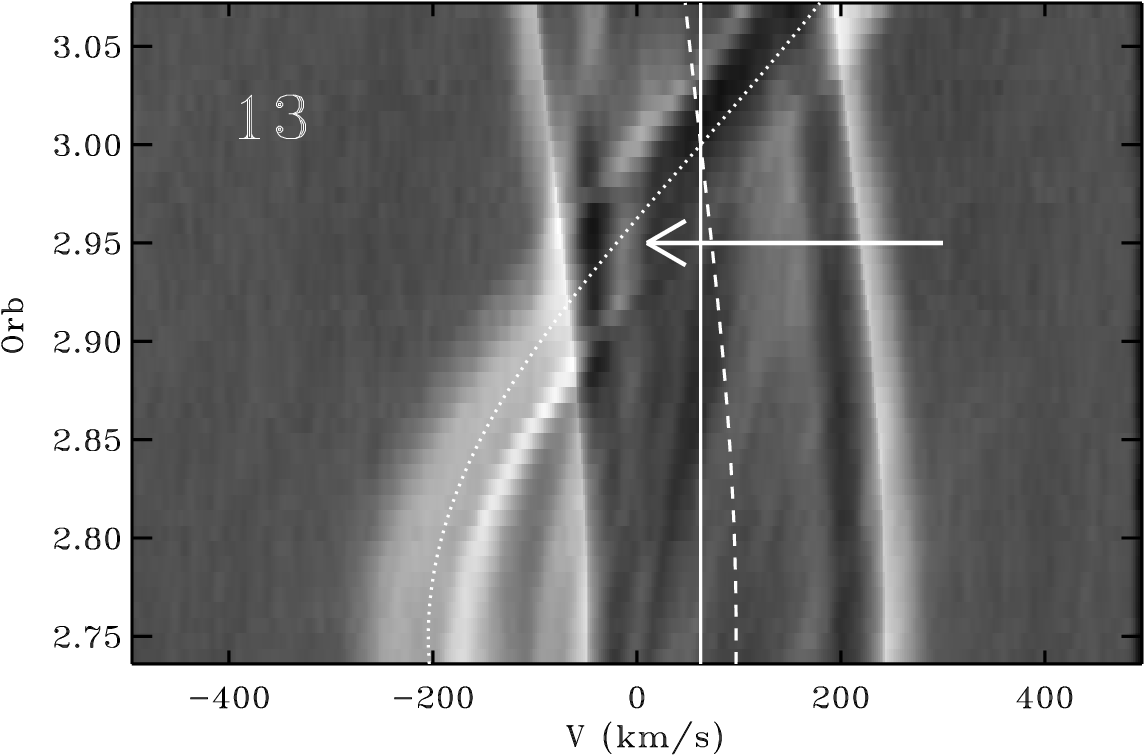}     %  image of wisp evol. in pri eclipse
\caption{ 
\footnotesize
The orbital-phase evolution of the Enhanced Spectral-line Perturbation (ESP) on
the secondary component of $\epsilon\,$CrA during the transit eclipse. 
The horizontal axis shows the radial velocity in the heliocentric system
while the vertical axis gives the time (increasing upwards) 
expressed in orbital periods, $Orb$; the eclipse centre on 
the observing night \#13 was located at $Orb = 3.0$ (see P2).
The ESP localized at the sub-observer point at $\phi \simeq 0.65$ (as described 
in the text and in Figure~\ref{fig_Eps_spot}), possibly continued its evolution into the eclipse
phases and merged with other spectral-line intensity features. 
The  mass-center velocities of both stars 
are shown by broken (primary) and dotted (secondary) sine-curve lines.
}
\label{fig_Eps_jump}
\end{center}
\end{figure}
% ------------------------------------------------------------------------------------------
 
A detailed analysis of the images formed from the BF profiles for $\epsilon\,$CrA
indicated an additional feature in the 
secondary-component profile that was not discussed in P2: 
The hot gas that produced
 the ESP remained at least partially visible during the transit-eclipse phases. 
As shown in Figure~\ref{fig_Eps_jump}, after encountering the primary profile
at $\phi \simeq 0.89 - 0.90$, the ESP stopped shifting in heliocentric 
radial velocities and stabilized at about $-75$ km~s$^{-1}$ 
relative to the binary mass center. This orientation corresponded to
the rapid variation of its RV relative to the secondary mass center.
All of this happens in projection over the large primary-component disk
 (Figure~\ref{fig_Eps_spot}). 
Because of the relatively low orbital inclination of $\epsilon\,$CrA of 
$i \simeq 72^\circ - 73^\circ$, it is possible that -- thanks to the line of sight 
passing {\it above\/} the secondary -- we see here
the volume between the binary components. This orientation can be visualized 
for transit-eclipse phases $\phi > 0.90$ in Figure~\ref{img_EpsCrA_Lucy}. 
From about $\phi \simeq 0.98$ onward, just before the mid-eclipse,
the active region became visible at negative velocities (Figure~\ref{fig_Eps_jump}),
presumably on the observer-approaching side of the secondary component. 
This may be produced by the fresh, hot gas from the primary sliding along 
the surface of the secondary component. 

% =====  end   Sect: Secondary Eps CrA =====================================

% ======= start == AW UMa Secondary and RV orbit  ====================    Sec.4

\bigskip

\section{The secondary component of AW~UMa} 
\label{sec-aw}

The most extensive previous spectroscopic investigation of the AW~UMa orbit
by \citet{PR2008} was conducted at a medium spectral resolution with data collected
at the David Dunlap Observatory (DDO) 
over ten non-consecutive nights. The mean secondary component 
profile was used to derive the semi-amplitude $K_2$ and then the
mass ratio, $q_{\rm sp} = 0.101 \pm 0.006$. 
This value was larger than any of the previous Lucy model 
photometric determinations, which tended to concentrate
around  $q_{\rm ph} \simeq 0.08$.

%  Figure: --------- AW UMa two edges of the secondary profile  defined ----------  Fig.5
\begin{figure}[h]                       % top, bottom, here
\begin{center}
\includegraphics[width=14.5cm]{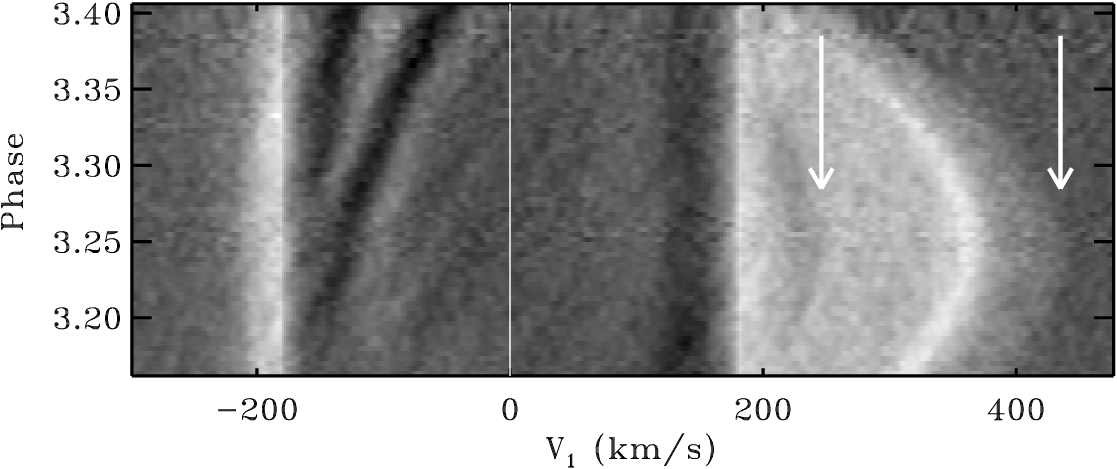}    %  image of AW UMa sec prof. 0.25 phase
\caption{
\footnotesize 
A  segment of the 2D image formed from individual BF's 
arranged in orbital phase for the second of the available three observing nights of
AW~UMa; it is a part of Figure~9 in P1. 
The radial velocities (the horizontal axis) are in the system centered on the primary
component with the strong rotational profile of the primary 
component subtracted from the BF profiles for better visibility of the secondary component.
 The orbital phases (the vertical axis), are expressed as a full
count of orbital cycles from the assumed initial epoch, as in P1.  
The edges of the secondary component 
profile, as marked  by the arrows, $E_1$ and $E_2$, were determined from
images similar to this one. 
% for phases free of the component-profile overlap.  
Note that the strongly non-uniform distribution of intensity across 
the secondary-component profile and the presence of the bright RV ``wisp''
which is identified in Figure~\ref{fig_AWUMa_spot} as ESP \#4. 
}
\label{fig_sec_edge}
\end{center}
\end{figure}
%  ------------------------------------------------------------------------------------

With only three available nights and given a highly complex 
secondary-component profile, the P1 analysis could provide only a tentative 
confirmation of the $q_{\rm sp} \simeq 0.10$ result. 
We re-discuss the results of P1 in light of the experience gained in 
analyzing the simpler and better-observed  $\epsilon\,$CrA binary
(Section~\ref{sec-eps}). It should be taken into account that 
only five orbital revolutions over three nights -- with interruptions -- were available  
for AW~UMa. As for $\epsilon\,$CrA,
a search was made for (1)~distinct RV edges of the secondary component profile 
(as for a detached star) that (2)~would move in anti-phase to the motion of the
primary component of the binary. They have been detected, but they are 
less distinct than in  $\epsilon\,$CrA due to the presence of several variable ESPs
(see Figures 8 -- 10 of P1). 
An example of the edges is shown in Figure~\ref{fig_sec_edge} where they 
are marked by arrows. 
The heliocentric velocities of the edges are listed in Table~\ref{tab_sec_aw}. 

%\bigskip

% Table:  AW UMa: Velocities of the secondary component, heliocentric ASCII ======== Table 1
%                                                                                   tab_sec_aw_helio.txt       [orb, E1, E2]
\begin{deluxetable}{CCC}[h]

\tabletypesize{\footnotesize}                 % 10 pts
\tablewidth{0pt}
\tablecaption{AW~UMa: Radial velocities of the secondary-component profile edges
\label{tab_sec_aw}}

\tablehead{
 \colhead{Phase ($Orb$)} & \colhead{$E_1$ (km~s$^{-1}$)} & \colhead{$E_2$ (km~s$^{-1}$)}
}
\startdata
   0.604  &  -303.33  &  -178.28 \\
   0.632  &  -336.09  &  -206.07 \\
   0.636  &  -334.82  &  -213.09 \\
   0.674  &  -379.83  &  -237.40 \\
   0.681  &  -377.60  &  -243.44
\enddata

\tablecomments{
The first column gives the time expressed in orbital phase 
counted continuously from the assumed initial epoch 
(the $Orb$ units, as defined in P1). \\
The second and third column give the heliocentric 
radial velocities of the edges of the
secondary component profile, $E_1$ and $E_2$ expressed in km~s$^{-1}$, 
as explained in the text.  \\
This table is available in the on-line version only. 
}

\end{deluxetable}
% ----------------------------------------------------------------------------------------

% Table:  ======= Orbital solutions  ========================================= Table 2
\begin{deluxetable}{CCCCl}[h]

\tabletypesize{\footnotesize}                 % 10 pts
\tablewidth{0pt}
\tablecaption{AW~UMa: Orbital solutions
\label{tab_orb}}

\tablehead{ 
\colhead{\#} &  \colhead{$V_0$}  &   \colhead{$K_1$}  &   \colhead{$K_2$}  &  \colhead{Comment} \\  
\colhead{}  &     \colhead{km~s$^{-1}$} & \colhead{km~s$^{-1}$} & \colhead{km~s$^{-1}$} & \colhead{}
}

\startdata
1 &  -15.90 \pm 0.12   &  28.37 \pm 0.37   &                             &  From P1, $0.25 < \phi < 0.75$, 299 obs. \\
2 &  -15.88 \pm 0.06   &  29.08 \pm 0.09   &                             &  same for $0.20 < \phi < 0.80$, 340 obs.  \\
3 &  -14.57 \pm 1.13   &                             & 314.06 \pm 1.18  & $0.15 < \phi < 0.35$ and $0.65 < \phi < 0.85$, 45 obs. \\
4 &  -15.74 \pm 0.14   &  29.08 \pm 0.09   & 314.32 \pm 1.17  & Combined solution, phases as for \#2 and \#3. \\
\enddata

\tablecomments{
The solution uncertainties were estimated as standard ($\pm 1\sigma$)  errors 
by a bootstrap-sampling experiment based on 10,000 repetitions. 
The bootstrap parameter distributions did not show any obvious asymmetries.
%The total number of available observations was 584 measurements  the primary (P1)
%and 118 measurements of the secondary. 
}

\end{deluxetable}
% ==================================================================

%  Figure:  ======== RV orbit for both components of AW UMa ====================  Fig.6
\begin{figure}[h]
\begin{center}
\includegraphics[width=12.5cm]{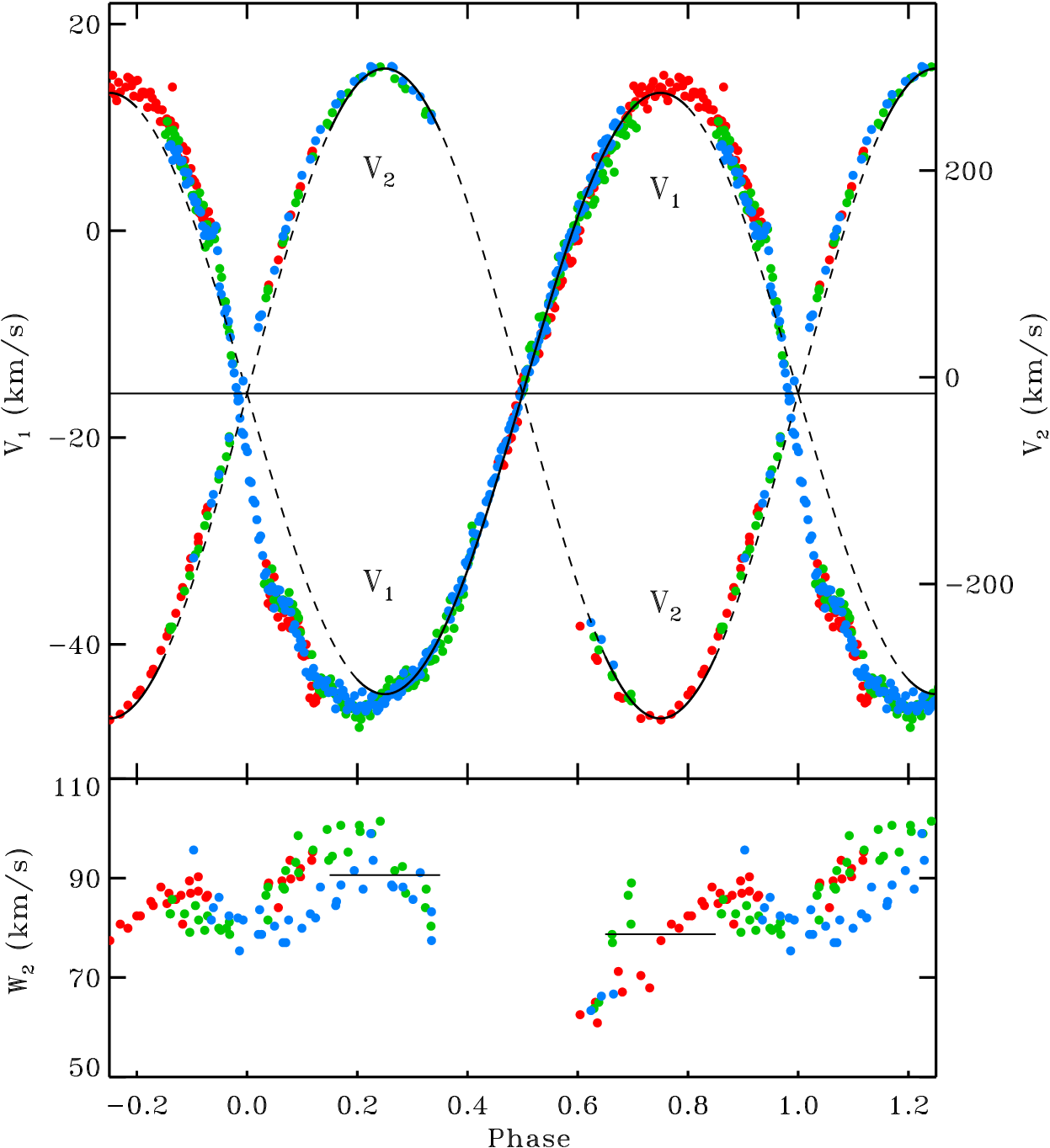}        % RV's AW UMa both components
\caption{
\footnotesize
The heliocentric orbital velocities of both AW~UMa components. 
The figure corresponds to Figure~4 in P1, 
with added velocities of the secondary component, derived
as the mean values of the secondary profile edges $E_1$ and $E_2$.
The scale for the secondary (the right vertical axis)  is compressed 10 times relative to that
for the primary and is shifted to the same $V_0$. 
The three observational nights of AW~UMa are coded by color: 1 -- red, 2 -- green, 3 -- blue.
The continuous lines show the sine curve fits to the data; the dashed line parts
correspond to the phase intervals which were not used in the fits because of mutual
interference of the wide component profiles.
The lower panel shows the half-width of the secondary component profile, 
$W_2 = (E_2 - E_1)/2$, which for a standard rotational profile  
would correspond to the apparent equatorial velocity $V_2 \sin i$. 
}
\label{fig_RVorbit}
\end{center}
\end{figure}
% =======================================================================

% ==============  Figure:   drift of the RV feature in AW UMa  =====================  Fig.7
\begin{figure}[ht]
\begin{center}
\includegraphics[width=12.5cm]{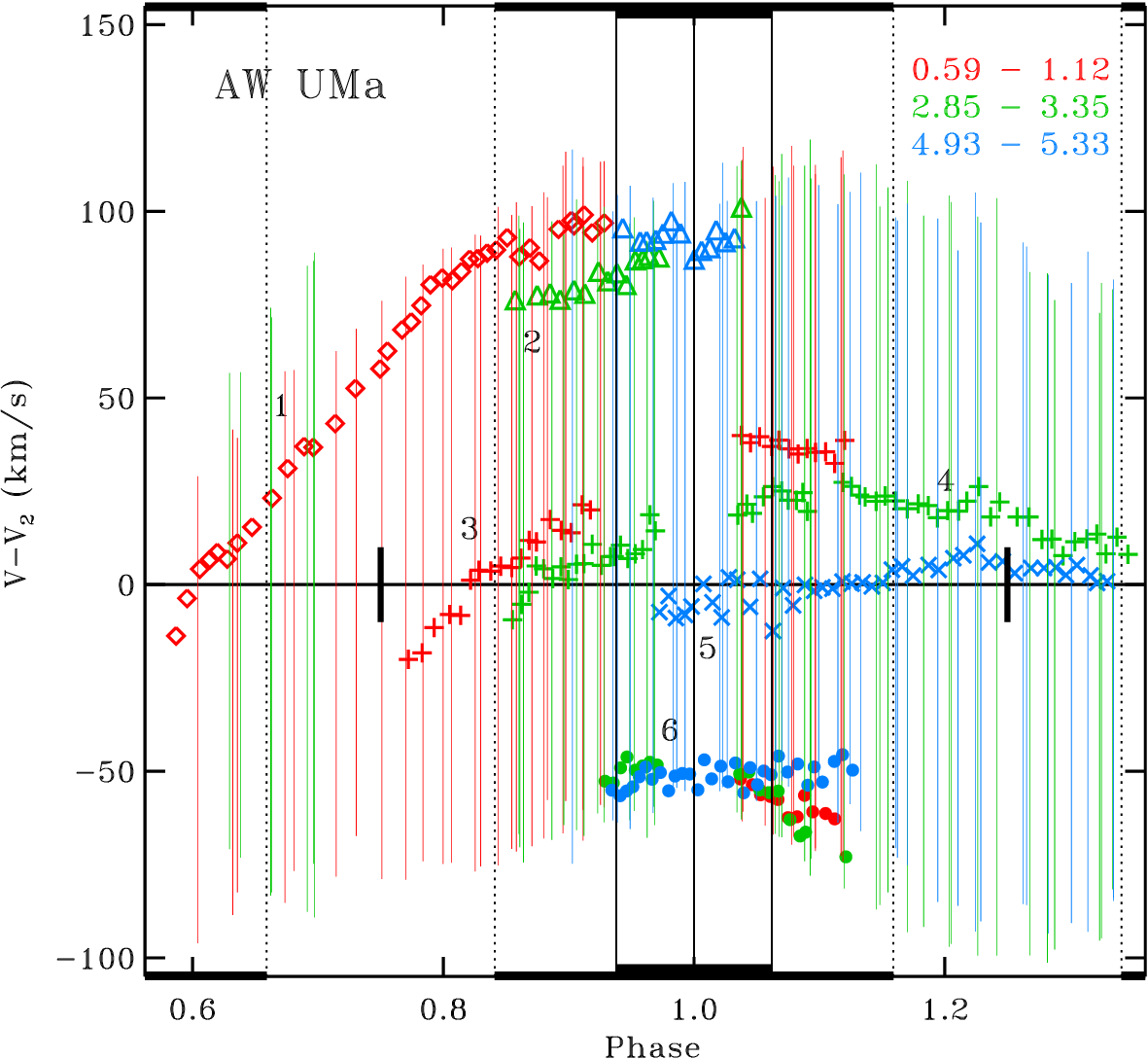}           %  RV ranges sec. AW UMa
\caption{
\footnotesize
The figure shows the AW~UMa secondary-component measured velocity extent 
as vertical lines coded in color by the phase range of
the observing night;  see the legend in the right upper corner of the figure. 
The figure is similar to the one for $\epsilon\,$CrA (Figure~\ref{fig_Eps_spot})
where other details are explained.  The numbered Enhanced Spectral-line Perturbations
(ESPs) observed on individual nights are discussed in the text. 
}
\label{fig_AWUMa_spot}
\end{center}
\end{figure}
% ========================================================================

% ======================================================================== Table 3
% Table:  Orbit and parameters of AW UMa

\begin{deluxetable}{cCCc}[h]

\tabletypesize{\footnotesize}                 % 10 pts
\tablewidth{0pt}
\tablecaption{The orbital parameters of AW~UMa
\label{tab_els_aw}}
%\tablenum{3}

\tablehead{
\colhead{Parameter } & \colhead{Result} & \colhead{$\pm\sigma$} & \colhead{Unit} 
}
\startdata
%                     res                         err                 units
 $P$             &   0.4387242       &                  &   day                 \\  % remove these 2 lines? 
 $HJD(pri)$  &  2455631.6498  &                  &   day         \\  % same as in P1
 $V_0$         &   -15.74 \phn \phn   &    0.14  &    km~s$^{-1}$   \\
$K_1$          &   29.08              &    0.09  &    km~s$^{-1}$   \\
$K_2$          &   314.32 \phn        &    1.17  &   km~s$^{-1}$    \\
\hline 
$q_{\rm sp} = M_2/M_1$ &   0.092    &   0.007    &                   \\
$A \sin i$     &  2.977             &   0.001    &   $R_\odot$        \\
$(M_1+M_2) \sin^3 i$ & 1.841        &   0.019    &   $M_\odot$        \\
$M_1 \sin^3 i$     & 1.685          &  0.020     &  $M_\odot$    \\
$M_2 \sin^3 i$     & 0.156          &  0.0026    &  $M_\odot$   % \\
\enddata

\tablecomments{
The errors are the formal least-squares errors and do not reflect possible
systematic uncertainties in the data. 
}

\end{deluxetable}
% ========================================================================

With the new RV data for the secondary component, we attempted a determination
of the spectroscopic orbital elements for AW~UMa.
Of the available 584 radial velocities for the primary component
and 118 measurements for the secondary, 
the orbital solution used  340 velocities within $0.2 < \phi < 0.8$
as listed in Table~3 of P1 and  all 118 measurements of the profile
edges $E_1$ and $E_2$, as listed in Table~\ref{tab_sec_aw}.
The mean values from the edges were used to represent 
the motion of the secondary component. 
% Although the secondary was  visible through the 
%transit eclipse phases, only the orbital-quadrature data were used within 
%the phases 0.15 -- 0.35 and  0.65 -- 0.85 in the solution to avoid mutual overlap
%of the profiles. 

The orbital solutions for AW~UMa are given in Table~\ref{tab_orb}. The first 
two lines quote the results for the primary component from P1, while the last line
gives the combined final solution for  the three
parameters $V_0$, $K_1$ and $K_2$.
The solution is well-defined thanks to the simultaneous constraint on $V_0$ 
from the motion of both components and the RV's of the
secondary component following the sine curve.
%\footnote{We expected that the circular motions is the only permissible
%one for highly dissipative, gaseous bodies such as stars in a contact binary.}. 
The uncertainties of the RV parameters $V_0$, $K_1$ and $K_2$ 
in Table~\ref{tab_orb} were determined by bootstrap-sampling  
experiment.  %, as commented in the table. 
While the formal random errors
estimated this way appear to be small, systematic errors may be larger
because of the lack of a proper velocity-field description on the stellar disks. 

The radial velocity variations for both components are shown 
in Figure~\ref{fig_RVorbit}. It is similar to Figure~4 in P1, but with the 
secondary-component orbit added using the 
velocity scale reduced by 10 times relative to that for the primary component. 
Table~\ref{tab_els_aw} gives the derived parameters of the binary.
Please note that the value of $K_2$ in Sec.~7.1 of P1 was incorrectly 
given to the sum of the semi-amplitudes; the correct numbers are given here. 

The new radial-velocity solution, as shown in Figure~\ref{fig_RVorbit},
confirms the previously noticed negative deviations of 
the primary-component radial velocities after the transit mid-eclipse,
in the wide phase range $0 < \phi <  0.25$.
Contrary to what was suggested in P1, the RV deviations do not appear to result
from the ``Rositter -- McLaughlin" effect since we see 
only {\it a trace of positive deviations before\/} the mid-eclipse. Thus, the 
deviations are asymmetric relative to the line joining the mass centers. The
change is forced mostly by the shift in $V_0$, caused by the inclusion
of the secondary orbit. 
The phases immediately after the mid-eclipse 
correspond to the orientation of the binary 
with the Coriolis-deflected flow from the primary directed at the observer. 

The negative RV deviations reach about $-10$ km~s$^{-1}$ 
and, surprisingly, the whole, broad profile has acquired such a negative shift
as if the whole hemisphere of the primary were approaching the observer. 
There exists a corresponding negative deviation after the mid-eclipse in
$\epsilon\,$CrA (see P2, Fig.4), but it is smaller, on the order of 
 $-3$ km~s$^{-1}$ and much more localized in the phase range. 
The current restriction of the AW~UMa primary data 
to the phase range $0.20 < \phi <  0.80$  
is an attempt to eliminate those (poorly understood) profile
distortions and thus restore symmetry relative to the line joining 
the two mass centers. 

We note that the primary profiles were observed 
to briefly broaden during the primary eclipse
phases in both binaries (see Fig.~3 in P1 and Fig.~3 in P2), 
but the width variations seemed to be symmetric relative to the primary
eclipse mid-eclipse phase. The findings of the
mean velocity shift and the broadening symmetry may be important
in understanding the mass outflow from the primary component. 

The radial velocities of the secondary component of AW~UMa, 
as derived from the two profile edges $E_1$ and $E_2$, generally follow
the anti-phase variations relative to the primary motion, but with
the similar irregularities to those observed in $\epsilon\,$CrA: 
\begin{enumerate}
%(1)~
\item The half-widths, $W_2 = (E_2 - E_1)/2$, are systematically variable:
At the first quadrature $W_2 (0.25) = 91 \pm 3$ km~s$^{-1}$ while at the second 
quadrature $W_2 (0.75) = 79 \pm 3$ km~s$^{-1}$ 
(the lower panel of Figure~\ref{fig_RVorbit});  
%(2)~
\item The mean velocities 
show $\pm 30$  km~s$^{-1}$  deviations from the expected sine shape
 (Figure~\ref{fig_AWUMa_spot}); they can be considered moderate 
 relative to the large semi-amplitude $K_2 = 314$  km~s$^{-1}$, but
 are systematic. 
 \end{enumerate}
 These irregularities, similar to those observed  in $\epsilon\,$CrA
(Section~\ref{sec-eps}, Figure~\ref{fig_Eps_spot}) probably reflect
changes in the spectral-line optical depth     % used in our work 
for the rapidly changing geometry of the binary system.

\medskip
While the extent and motion of the underlying secondary-component profiles
in AW~UMa and $\epsilon\,$CrA appear to be similar, the
two binaries differ in the type and number of the ESPs.
One strong ESP was visible for $\epsilon\,$CrA on all nights (Figure~\ref{fig_Eps_spot});
it always showed the same phase drift as the binary rotated. 
In contrast, in AW~UMa two to four ESPs were simultaneously visible
(Figure~\ref{fig_AWUMa_spot}) at different locations within the secondary component
profile. One of these was very similar to that in
$\epsilon\,$CrA, while the others appeared more irregular showing migration
within hours and sometimes reappearing after the observing daytime breaks.
The ESPs were typically stronger than the underlying profile;
an example, ESP\#4,  is visible in Figure~\ref{fig_sec_edge}. 
% projecting on the secondary profile.

The very short period of AW~UMa created another complication in attempts
to follow the evolution of individual ESPs:
expressed in the binary orbital cycles, the CFHT observations experienced relatively
long daily breaks amounting to about 1.5 orbital periods of the binary. 
The observing run was short, lasting only three consecutive nights
or five binary revolutions. 

Figure~\ref{fig_AWUMa_spot} shows the ESP phase locations and phase
evolution in AW~UMa.
The apparent equivalent of the single, well-defined feature in 
$\epsilon\,$CrA, ESP\#1, could be observed only on the first
night at a similar sub-observer phase $\simeq 0.6 - 0.65$, as for $\epsilon\,$CrA.
It is possible that ESP\#2 was a continuation of \#1 on the 
two following nights; its velocities corresponded to the rotationally 
receding side of the secondary component.  
The features \#3, \#4 and \#5 appeared at velocities similar to
the presumed velocity of the secondary mass center. 
They tended to drift slightly on all three nights, with the
region \#5 being most stable at the mean velocity of the secondary. 
Finally, the repeatable and well-defined ESP\#6 was observed at negative 
velocities; it was most probably produced by the matter 
moving along the surface of the secondary on its side approaching the observer.
% practically identically as the similar feature in  $\epsilon\,$CrA. 

While none of the ESPs  can be explained within the Lucy model,
the St\c{e}pie\'{n} model actually {\it predicts\/} the existence of features 
such as those given numbers \#1 and \#6. %for AW~UMa.
The ESPs would be manifestations of the stream encompassing 
the binary system; the sufficiently spectral-line optically thick layers of the
stream are visible emerging from behind the secondary component during
the transit phases when they point at the observer and are visible as ESP\#6.   
The stream continues around the secondary component
and eventually strikes the outer layers of the primary, sending the hot gas into the
overlying space; this region is visible as ESP\#1 in AW~UMa and as the
corresponding single ESP in  $\epsilon\,$CrA. However, the ESP features
close to the center of the AW~UMa secondary profile do not have an explanation.
%We return to the subject of the contact binary model in Section~\ref{model}. 

% === End =====  Section: AW UMa secondary ==================================

% ========= Section   mass-ratio discrepancy ==============================  Sec.5
\bigskip

\section{The mass-ratio discrepancy}
\label{discr}

% paragraph rewritten
There is a discrepancy in the mass-ratio determinations for AW~UMa, with
 spectroscopic results ($q_{\rm sp}$) consistently being larger than 
 photometric results ($q_{\rm ph}$). 
The new spectroscopic determination, as listed in Table~\ref{tab_els_aw},
$q_{\rm sp} = K_1/K_2 = 0.092 \pm 0.007$, is close to the previous
David Dunlap Observatory medium-resolution determination by \citet{PR2008}, 
$q_{\rm sp} = 0.101 \pm 0.006$. 
The difference between the two determinations most likely
reflects not only the use of different spectral resolutions
but also the very different temporal coverage during the two observing programs;
the DDO result utilized 10 partial nights distributed over
a longer interval in contrast to the short, intense, hight-resolution run.
%It should be stressed that 
Both spectral determinations are larger than several % Lucy-model
photometric determinations by different authors (see references in P1,
particularly \citet{Wilson2008}), some claiming errors as small as 0.0005
tending to concentrate around $q_{\rm ph} = 0.080 \pm 0.005$.
%All photometric determinations utilized the Lucy model.  
%As we discuss further in the paper, we are very sceptical about 
%photometric determinations of $q_{\rm ph}$ utilizing
%the Roche/Lucy model for W~UMa binaries, both due to the instability
%of the determination process (Section~\ref{mod-q}) and the physical viability
%of that model (Section~\ref{model}).
  
The existence of a possible  $q_{\rm ph}$ vs.\  $q_{\rm sp}$ discrepancy
in AW~UMa 
is important because the mass ratio is a fundamental parameter of a close
binary system and plays a special role in the Lucy model in
setting the value of the common equipotential. 
Indications of a possible discrepancy were signalled as early as  
\citet{ND1991}. At that time, spectroscopic                       %$q_{\rm sp}$ 
determinations for W~UMa binaries were lagging in both quantity and quality behind 
photometric Lucy-model determinations, as 
many spectral efforts were reduced to {\it spectroscopic detection\/}  of 
faint secondary-component signatures. Some results appearing in the literature still 
used the inappropriate -- for W~UMa binaries -- spectral 
``line-by-line'' radial-velocity measurements. 
 \citet{ND1991}  correctly explained the mass-ratio discrepancies
 as mainly due to poor accuracies of $q_{\rm sp}$ determinations of that time.
 
 AW~UMa was one of the first objects used to test modern spectroscopic techniques 
that utilized combined RV information from many spectral lines. This was mainly
 due to its brightness but also because of its unexpectedly small mass ratio 
 indicated by the light-curve photometric solutions, $q_{\rm ph} \simeq 0.08$.
% Initially, the new spectroscopic approaches 
First attempts utilized the inherently non-linear
 Cross-Correlation Function (CCF) technique applied  to also non-linear 
 photographic data \citep{McL1981,And1983,Rens1985}.
 These were later replaced by 
 %tests of 
 a linear deconvolution (Broadening Function) technique 
 applied to digital results  \citep{Rci1992}.
% \citep{Rci2002b}.
Because of low detector sensitivity, limited large telescopes
 time, and the unexpectedly weak signature of the secondary component 
 in AW~UMa, early analyses     %eg.\  \citet{Rci1992}  
 were limited to confirming the photometric mass ratio. 
The first spectroscopic determination 
%specifically avoiding any prior assumptions and 
based on more extensive material by  \citet{PR2008},
 $q_{\rm sp} = 0.101 \pm 0.006$, 
 clearly demonstrated that the mass ratio must be larger than previously thought. 
% Interestingly, 
This result coincided with a theoretical analysis of  \citet{BeP2007}, which pointed out 
 difficulties in explaining AW~UMa as the result of the 
 close binary-star evolution for $q < 0.10$. 
 
 % Table:   ---------- Comparison q(sp) vs. q(ph)   the sample  ------------------
%                                                                            
\begin{deluxetable}{lccccccc}[t]

\tabletypesize{\footnotesize}                 % 10 pts
\tablewidth{0pt}
\tablecaption{The $q_{\rm ph}$ and $q_{\rm sp}$ determinations of totally eclipsing systems}
\label{tab_Qtab}

\tablehead{
 \colhead{Name} & \colhead{$P(d)$} & \colhead{$B-V$}    & 
     \colhead{$q_{\rm ph}$}  & \colhead{$\sigma q_{\rm ph}$}  &
     \colhead{$q_{\rm sp}$}  & \colhead{$\sigma q_{\rm sp}$}   & Refs.  }
 %     Name       Per    B-V     q(ph)    err    q(sp)    err   Ref
 
\startdata
 \sidehead{$q_{\rm sp}$ from the DDO survey}
 CC Com    & 0.2207 & 1.240  & 0.521 & 0.004 & 0.527 & 0.006 & (1, 2)  \\         
 V1191 Cyg & 0.3134 & 0.390 & 0.094 & 0.005 & 0.107 & 0.005 &  (3, 4) \\         
FG Hya       & 0.3278 & 0.540 & 0.138 & 0.009 & 0.112 & 0.004 & (5,6,7,  8) \\  
BB Peg       & 0.3615 & 0.480 & 0.356 & 0.003 & 0.360 & 0.009 & (9,  8)  \\         
AM Leo       & 0.3658 & 0.490 & 0.398 & 0.003 & 0.459 & 0.004 & (10,  2) \\         
V417 Aql     & 0.3703 & 0.570 & 0.368 & 0.001 & 0.362 & 0.007 &  (11, 8) \\  
HV Aqr        & 0.3745 & 0.470 & 0.146 & 0.020 & 0.145 & 0.050 &  (12, 13) \\       % q(sp) large err.
V566 Oph   & 0.4096 & 0.410 & 0.238 & 0.005 & 0.263 & 0.012 & (5, 14)   \\
Y Sex          & 0.4198 & 0.390 & 0.175 & 0.002 & 0.195 & 0.010 & (15, 16) \\ 
EF Dra        & 0.4240 & 0.460 & 0.125 & 0.025 & 0.160 & 0.014 & (17, 8) \\
AP Leo        & 0.4304 & 0.470 & 0.301 & 0.005 & 0.297 & 0.012 & (18, 8) \\
AW UMa     & 0.4387 & 0.360 & 0.080 & 0.005 & 0.092 & 0.007 &  (19,20, 21,22)  \\  % not in the DDO survey
DK Cyg       & 0.4707 & 0.380 & 0.330 & 0.020 & 0.325 & 0.040 &  (5, 23)   \\
UZ Leo        & 0.6180 & 0.350 & 0.230 & 0.003 & 0.303 & 0.024 & (24, 23) \\
 \sidehead{$q_{\rm sp}$ from other sources}    %    \citet{ND1991}
RZ Com      & 0.3385 & 0.700 & 0.420 & 0.010 & 0.430 & 0.030 & (25, 26) \\  % modified q(ph)
AE Phe       & 0.3624 & 0.640 & 0.391 & 0.005 & 0.450 & 0.010 & (27, 28) \\
AQ Tuc       & 0.5948 & 0.400 & 0.265 & 0.010 & 0.350 & 0.010 & (29, 29) \\ % full discussion of discrep. 
RR Cen      & 0.6057 & 0.360 & 0.180 & 0.010 & 0.210 & 0.010  & (5, 30) \\
MW Pav     & 0.7950 & 0.33   &  0.182 & 0.003 & 0.228 & 0.006  & (31, 32) \\         %   DR1
 \enddata

\tablecomments{
References to individual mass-ratio determinations ($q_{\rm ph}$, $q_{\rm sp}$): \\
1.~\citet{Rci1976}, 2.~\citet{DDO-12},            % CC Com,      DDO-12
3.~\citet{Prib2005}, 4.~\citet{DDO-13},           % V1191 Cyg,  DDO-13
5.~\citet{MD1972b},  6.~\citet{Tw1979},   7.~\citet{Yang1991},  8.~\citet{DDO-1},       % FG Hya
%9.~\citet{CS1981},       % BB Peg not used, 
9.~\citet{Leung1985},      % BB Peg 
10.~\citet{Hiller2004},      % AM Leo
11.~\citet{Samec1997},   % V417 Aql
12.~\citet{Robb1992},  13.~\citet{DDO-3},    % HV Aqr
14.~\citet{DDO-11},      %   V566 Oph
15.~\citet{Hill1979}, 16.~\citet{DDO-15},     % Y Sex
17.~\citet{Plewa1991},        %   EF Dra
18.~\citet{Zhang1992},
19.~\citet{MD1972a}, 20.~\citet{WVH2009},  
          21.~\citet{PR2008},  22.~\citet{Rci2015}[P1],    % AW UMa
23.~\citet{DDO-2},                                                  % DK Cyg
24.~\citet{Vinko1996},                                             % UZ Leo
25.~\citet{WD1973},                  %  RZ Com ph   Wils & Dev. 1973mean from 2 light curves 
26.~\citet{MH1983},                  % RZ Com sp   McLean & Hild 1983
27.~\citet{ND1991},                % AE Phe ph several sources, q(ph)  range:  0.384 - 0.396 
28.~\citet{Duer1978},                % AE Phe sp
29.~\citet{HK1986},                  % AQ Tuc ph, sp
30.~\citet{KH1984},                  % RR Cen sp
31.~\citet{Lapa1980},                % MW Pav ph
32.~\citet{DR1}                      % MW Pav sp
}

% MW Pav:    Lapasset, E. 1980, \aj, 85, 1098       {Lapa1980}           %  MW Pav
% DR1:     Rucinski, S. M. \& Duerbeck, H. W. 2006, \aj, 132, 1539       % DR1
% DR2:      Duerbeck, H. W. \& Rucinski, S. M. 2007, \aj, 133, 168        % DR2

\end{deluxetable}
% -------------------------------------------------------------------------------------------

%While the  $q_{\rm sp}$ vs.\ $q_{\rm ph}$ discrepancy appears to
% be well established for AW~UMa, 
A similar discrepancy as in AW~UMa is very weakly evident for $\epsilon\,$CrA, though 
%where the values of $q_{\rm sp}$ and $q_{\rm ph}$ are very close, 
still in the same sense,       % as for AW~UMa, 
with $q_{\rm sp} = 0.1300 \pm 0.0010$ (P2) compared with
 $q_{\rm ph} = 0.114 \pm 0.003$   \citep{Tw1979} and
 $q_{\rm ph} = 0.1244 \pm 0.0014$  \citep{WR2011}.
$\epsilon\,$CrA may show partial eclipses so that its 
photometric determination of  $q_{\rm ph}$ may be
less reliable than for a totally eclipsing binary such as AW~UMa.
 
A comparison of $q_{\rm ph}$ with  $q_{\rm sp}$  utilizing 
much better data than those available to  \citet{ND1991} is now possible. 
A large body of spectroscopic determinations resulted from a 
program of RV observations of short-period ($P < 1$ day), bright binaries
conducted in years 1993 -- 2010 at the DDO
and published in 23 papers in years 1993 -- 2010. 
The program included 124 W~UMa-type binaries. The spectroscopic
determinations during the DDO program 
did not use any previous, literature
data by design, ensuring no bias from published $q_{\rm ph}$.
The binaries were selected for observations 
based   %simply 
on the sky accessibility, star brightness, 
and the somewhat erratic weather conditions at the DDO. 
A    %convenient 
list of the observed binaries is provided in \citet[Table 1]{Rci2013a}.

The size of the DDO spectroscopic program and the consistency of the 
methods used ensure uniformity of the $q_{\rm sp}$ determinations.
% and permits a discussion of the $q_{\rm ph}$ vs.\  $q_{\rm sp}$ discrepancy 
% in view of predictions of the currently available models. 
In contrast, the %$q_{\rm ph}$ 
photometric determinations come from various sources in the literature.
All researchers utilized the light-curve synthesis model developed
following the \citet{Lucy1968b} description, which was first 
extensively used by \citet{MD1972a,MD1972b}. The light-curve
synthesis approach gained widespread utilization after its implementation by  
\citet{WD1973}, followed by several widely accessible computer codes
such as \citet{WVH2009,Prsa2016,Conr2020,2020ascl.soft04004W}.

 % ------------------  Figure:     q(ph) - q(sp)  ----------------------------------  Fig.9
\begin{figure}[h]
\begin{center}
\includegraphics[width=9.5cm]{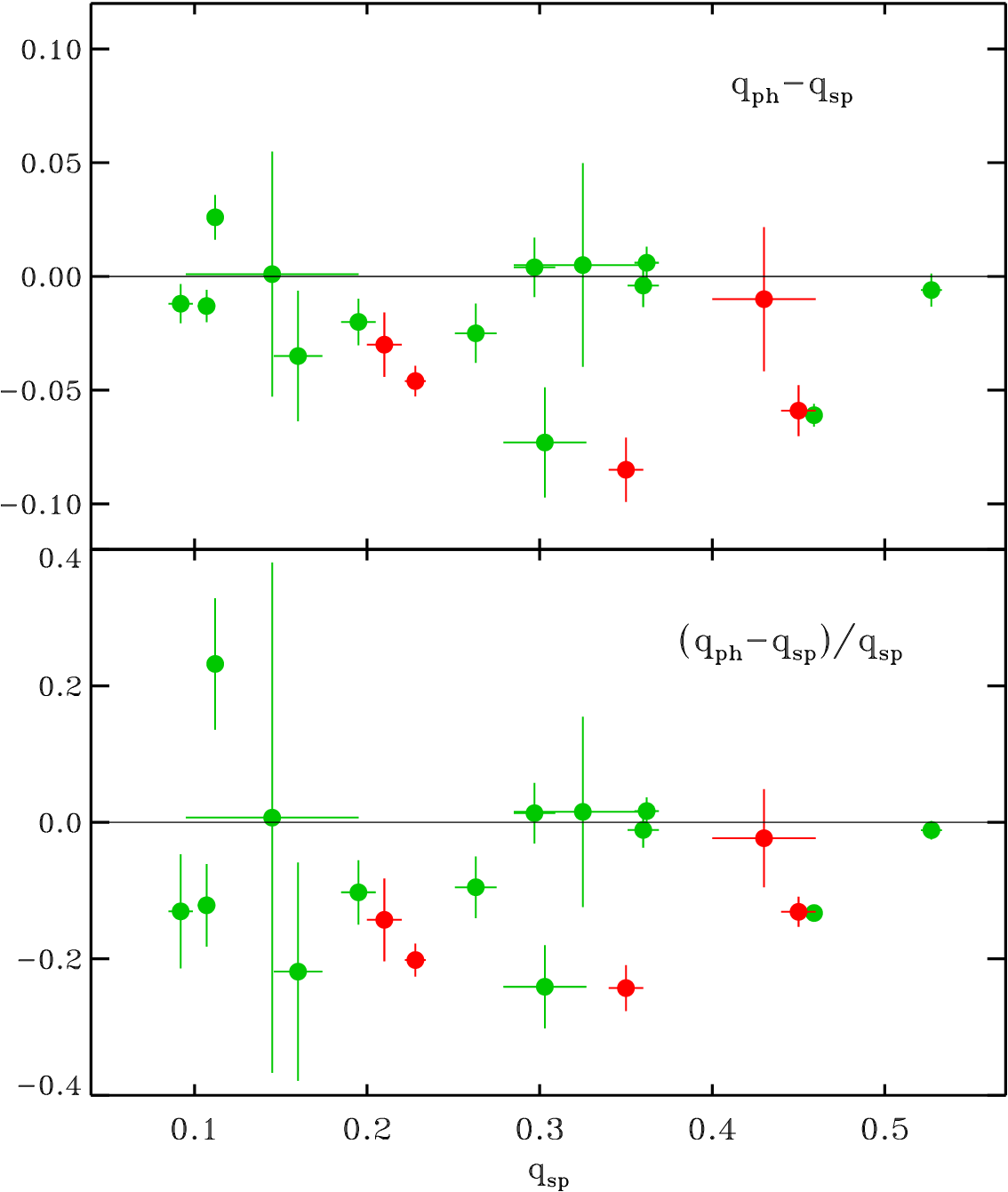}       %  mass-ratio discrepancy
\caption{
\footnotesize
The mass-ratio discrepancy between the photometric,  
$q_{\rm ph}$, and spectroscopic, $q_{\rm sp}$, determinations is shown 
directly versus the spectroscopic value (the upper panel) and
normalized by $q_{\rm sp}$ (the lower panel).
The data shown in green correspond to $q_{\rm sp}$ determinations from the
DDO survey while those in red are critically updated 
spectroscopic results listed by \citet{ND1991}. See the text for details.
}
\label{fig_q-discrep}
\end{center}
\end{figure}
% ------------------------------------------------------------------------------------------------

Certain precautions were applied to ensure the mutual independence of 
the  $q_{\rm sp}$ and $q_{\rm ph}$ determinations. The selection 
included only: 
(1)~W~UMa-type binaries with $q_{\rm ph}$ Lucy-model determinations 
published {\it before\/} the spectroscopic determinations, 
and (2)~binaries showing total eclipses to ensure the 
highest photometric solution stability irrespectively of the specific
implementation of the Lucy model.  
The first condition aimed to prevent a ``leak'' and stabilization of
the photometric, iterative, multi-parameter solution on the already available 
spectroscopic value, while the second condition 
excluded poorly-conditioned light curve solutions based only on distortion
effects. 
The selected binaries are listed in Table~\ref{tab_Qtab} which is similar to
Table~2 in  \citet{ND1991}. 

The binaries with $q_{\rm sp}$ from the DDO program are marked 
in green in Figure~\ref{fig_q-discrep} and listed in the first half of
the table. We added binaries fulfilling the above restrictions from other sources,
including the \citet{ND1991} list, but only those 
with $q_{\rm sp}$ values determined using the CCF or BF deconvolution 
methods by other researchers. These results are listed in the second half of  
Table~\ref{tab_Qtab} and are marked by red symbols in the figure. 
Note that $\epsilon\,$CrA is not included in the table since its primary 
eclipse may be partial or close to partial.  
%
%               the St\c{e}pie\'{n} model

The difference $(q_{\rm ph}-q_{\rm sp})$ is shown in 
Figure~\ref{fig_q-discrep} directly and, similarly as in  \citet{ND1991}, 
in the relative sense, divided by the value of $q_{\rm sp}$.  
The normalization is kept here to reflect our lack of knowledge on
the source of the discrepancy. 
%may artificially amplify discrepancies at the low-end of the mass ratio. 
Regardless of the presentation, we observe
a widespread occurrence of $q_{\rm ph} < q_{\rm sp}$, exactly as
for AW~UMa, which is the binary with the smallest $q_{\rm sp}$ in the figure.
The size of the discrepancy is moderate, so either 
$q_{\rm sp}$ or $q_{\rm ph}$ could serve as an argument in the figure. 
% here modified more appreciably   
The author believes that the spectroscopic $q_{\rm sp}$  results
are currently better-defined because of their more straightforward derivation.
We return to this subject in Section~\ref{model}. 

The only binary in Figure~\ref{fig_q-discrep} showing
the photometric value of $q_{\rm ph}$ larger than the spectroscopic $q_{\rm sp}$ 
is FG~Hya. 
This binary has an extensive and consistent record of several photometric
determinations, so it urgently requires a confirmation of the spectroscopic result.
We note the spectroscopic result was obtained at the 
very beginning of the DDO program \citep{DDO-1}.
% DDO-1 may be a poor result

% this section edited/shortened
%The study of \citet{ND1991} presented another strong indication
%that $q_{\rm ph}$ may be affected by phenomena occurring within the binary 
%system, potentially affecting the applicability of the photometric Lucy model.
%The authors were confronted with excellent light curves of AE~Phe that 
%displayed significant season-to-season changes. One light curve showed 
%a very well-defined {\it total\/} occultation eclipse, while the other light curves 
%showed partial eclipses, usually associated with extensive out-of-eclipse 
%light curve variability. This transition from the total to the partial eclipse
%already suggests a potential problem with the common-equipotential model for the binary. 
The study of \citet{ND1991} provided another strong indication
that $q_{\rm ph}$ may be affected by phenomena not accounted in the
photometric model. The light curve of AE~Phe showed  
significant season-to-season changes, with one light curve showing
a well-defined {\it total\/} occultation eclipse.
The authors modeled the four seasonal curves using the Lucy model and found 
 a range of $q_{\rm ph}$ results, from $0.3836 \pm 0.0014$ to $0.3964 \pm 0.0010$.
%Although the formal uncertainties were small, as is almost always the case with
%the Lucy model solutions, 
While the formal uncertainties are undoubtedly under-estimated,
a difference of more than 0.01 between 
individual determinations of $q_{\rm ph}$ seems to be real and is
highly concerning.   %Strictly speaking, 
Such a discrepancy would suggest a large and rapid mass 
exchange between the components, amounting to 
$0.01\,M_\odot$, yet without an accompanying 
significant orbital-period variation. 

The St\c{e}pie\'{n} model (Section~\ref{mod-kst}) offers a possible 
explanation for the transitions between rounded and flat-bottomed occultation
 eclipses in AE~Phe. These variations could result from changes in the overall 
 loading of the flow (also referred to as the ``equatorial bulge'') 
 around the secondary component. Unfortunately, no reliable $q_{\rm sp}$
determination is available for this star at present.
 
 % ===== end ==== Section:  Discrepancy in q ===================================

% ============= Section Methods Determination of q ======================    Sec.6

\bigskip

\section{The photometric and spectroscopic mass-ratio determinations}
\label{mod-q}

The tendency for $q_{\rm ph} < q_{\rm sp}$ is an important indication that one of the
methods used for mass-ratio determinations -- either the photometric Lucy model 
or the spectroscopic direct approach -- may produce systematically incorrect results. 
We note that the results of this study, along with those from our 
two previous papers, definitely confirm the existence of 
a velocity field in the rotating system of coordinates. 
Thus, strictly speaking, equipotentials cannot be defined. 
However, for low velocities, stellar shapes may still be close to predicted
to the equipotential geometry. We return to this subject in Section~\ref{model}. 

%This fact eliminates the concept of equipotentials, which simply {\it cannot be defined.\/}. 
%In Section~\ref{model}, we discuss the \citet{Step2009}      %{Step2013}
%model for W~UMa binaries, which explicitly includes velocities.
%This model is more complex than the Lucy 
 %model and necessitates abandoning the photometric determination of
 %$q_{\rm ph}$. Before that, we will first consider the currently used photometric
 % method of $q_{\rm ph}$ derivation, which does utilize equipotentials.
  
%The one-to-one dependence of the Roche geometry on the mass ratio 
%is a fundamental feature of the Lucy model and the foundation for its widespread 
%success in light-curve synthesis codes. 
%The dependence of the radius ratio, $r_2/r_1$, on the mass ratio, $q = M_2/M_1$, 
%is a strict relation for an equipotential surface selected 
%to represent the common envelope of the binary. We suggest that this assumption
%(1)~likely leads to larger uncertainties in $q_{\rm ph}$ than claimed, and 
%(2)~may indicate that the $q_{\rm ph}$ vs.\ $q_{\rm sp}$ 
%discrepancy highlights a failure of the Lucy model.  

The determination of photometric parameters for a W~UMa-type
binary is a complex process.
Initially, a $3D \rightarrow 1D$ mapping of the stellar {\it brightness\/}  
(but not velocities) % needed for probing the dynamics) 
is performed by the revolving binary itself, which
is registered as a light curve. This light curve is then subject to 
a multi-parameter solution utilizing the Roche-model geometry, 
with additional Lucy-model 
prescriptions for handling stellar-atmosphere properties. 
The solution is a highly non-linear, iterative process that includes 
unknowns such as the degree of contact (or the actual value
of the common equipotential),
and the orbital inclination, which are intermixed with assumed atmospheric
properties, such as the local flux level estimated by 
 surface brightness (effective temperature) 
and the limb and gravity darkening laws, typically derived from spherical
stars. Due to inherent nonlinearities,
this parameter determination process is highly susceptible to biased estimates
and may result in deceptively small uncertainties 
when estimated as if for a linear, correlation-free problem based on
the shape of the adopted deviations minimum.  
Unfortunately, this can lead to popularly quoted but
suspiciously small formal uncertainties. 

The complexity of  photometric determination of $q_{\rm ph}$ 
can contribute to the observed $q_{\rm ph}$ vs.\ $q_{\rm sp}$ discrepancy.
%To illustrate a prevailing nonlinearity, 
%let's specifically consider the size vs.\  mass ratio dependence, % $r = r(q)$,
%which is essential for the $q_{\rm ph}$ determination.  
Eclipse effects carry relatively  more information  
than the strongly model-dependent stellar-distortion effects. 
For spherical stars, it is known that the radius ratio, $k = r_2/r_1$
\citep{Russ1912}, is the key parameter 
controlling the depth and shape of eclipses.  
Although the binary components in our case are not spherical stars,
an equivalent of $k$ exists, albeit hidden within complexities of
light-curve-synthesis codes.
The individual dimensions of Roche common equipotential surfaces,
expressed, for example, as side equatorial radii $r_1$ and $r_2$, 
depend solely on the mass ratio, $r = r(q)$.  
This relation is known to be non-linear.
A simplified version of it has been extensively used by practitioners 
of close-binary evolutionary calculations to estimate the size of the 
Roche-lobe (inner critical equipotential) filling component. In
the orbital separation units, $r = 0.38 + 0.2 \log(q)$; this equation
works for both components (with inverted $q$)
and has an asymptotic low mass-ratio extension  \citep[Eq.4]{BeP1971}.  
By being approximately logarithmic,
the dependence $r = r(q)$ is a slowly-varying function. The weak dependence
of the Roche-lobe radii on $q$  is convenient for 
stellar binary models. However, $q_{\rm ph}$ determination utilizes 
an inverse relation, which -- consequently --  is a rapidly varying function. 
For the  approximation as above, it is an {\it exponential\/}
function, $q = 10^{5r-1.9}$. Therefore, small errors in the relative radius  
determination produce large deviations in the estimated mass ratio. 
An error of 1\% in $r$ produces a 4\% error in an $q_{\rm ph}$ estimate. 

The mass ratio is not determined from the sizes of the individual
Roche equipotentials, but from the ratio of their radii, $k = r_2/r_1$. 
This reduces the exponential nature of the relationship, making the non-linearity milder.
Using the same approximation for the inner critical Roche lobe 
as mentioned earlier, an approximate relation for the whole range $0 < k < 1$ remains
nonlinear, $q \simeq k^{+2.23 \pm 0.03}$. Therefore, 
a 1\% error in $k$ is expected to produce a 2.2\% error in $q_{\rm ph}$. 

The non-linearity in the $q_{\rm ph}$ determination,
as described above, has not yet been recognized.
%The use of different photometric codes and  specific assumptions
%could potentially diminish its influence; thus, {\it multiple independent\/}
%solutions utilizing different codes might somehow average out the 
%non-linearities. 
A more concerning issue is the {\it tendency\/} for 
$q_{\rm ph} < q_{\rm sp}$ among individual binaries (Section~\ref{discr}), with 
some binaries exhibiting this behavior while others do not.  
While we %strongly 
believe that an incorrect photometric model is the reason, some 
photometric solutions may provide systematically smaller values of 
$q_{\rm ph}$ due to the presence of an unrecognized 
``third light'' in the light curve. 
Companions to W~UMa binaries are very common, as demonstrated by
\citet{Toko2006} and \citet{Rci2007}. 
A light curve containing  unrecognized additional light has 
a reduced amplitude and may result in a $q_{\rm ph}$ value that is too small. 
That was the case for the photometric solution of one of our
two targets, $\epsilon\,$CrA, analyzed by \citet{ShZ2006}; they commented on 
difficulties in obtaining the best fit starting from the assumed (and correct)
spectroscopic value of $q_{\rm sp} = 0.129$ \citep{GD1993},  being
forced to reduce $q_{\rm ph}$ by 0.02.  $\epsilon\,$CrA  is not
listed in Table~\ref{tab_Qtab} precisely because it does not show
total eclipses and therefore cannot provide a secure radius-ratio determination.  
%These considerations do not seem to apply to AW~UMa where the 
%companion is relatively far away. 

The spectroscopic approach offers an entirely different and 
independent route of the mass ratio determination 
involving a $3D \rightarrow 2D$ mapping,
from the spatial velocities into a series of RV profiles arranged in time. 
This approach utilizes more accessible information in two dimensions of the
time and radial velocity; the RV data provide 
the orbital semi-amplitudes $K_1$ and $K_2$ leading to  
$q_{\rm sp} = K_1/K_2$. For very tight orbits, as observed for the
W~UMa-type binaries, with strongly dissipative, gaseous
bodies involved, the only permitted orbits are circular, so that the 
derivation of the semi-amplitudes is particularly simple.  
%Thus -- in principle -- the spectroscopic $q_{\rm sp}$ is obtained 
%with a lesser danger of biased results. 
%Here, the main uncertainties relate
%to the proper localization of the RV centroids and their coincidence with 
%the projected mass centers. Our results for 
%AW~UMa and $\epsilon\,$CrA in P1 and P2 show that the inter-binary velocity 
%fields are present with velocities reaching perhaps as much 
%as 30 -- 50 km~s$^{-1}$. These fields may affect the observed 
%velocity locations of the component mass centers. While the rotational-profile 
%centroids seem to provide very well defined orbital
%velocities of the primary components of both of our targets,
%the confined velocity extent of the secondaries -- similar to components 
%of detached binaries --  is an unexpected discovery and must be confirmed.
%It remains to be confirmed that centroids of such 
%very faint features correctly trace the motion of the secondary components. 

Although the direct approach using velocities suggests a lesser risk of biased 
results for the spectroscopic $q_{\rm sp}$, the values of the amplitudes
$K_i$ may still be biased for other reasons. 
Paradoxically,  the $K_2$ amplitude, which was very hard 
to determine in our binaries, seems to be more reliable, but only 
under strict conditions on the spectral feature used for the $K_2$ estimate.
The feature: 
(1)~must be free of any component profile overlap,
(2)~must move with the orbital phase in a sinusoidal manner, 
(3)~and strictly in anti-phase to the motion of the primary component. 
Obvioulsy, any inter-binary velocity features must be entirely avoided. 
%velocities that may reach as much as 30 -- 50 km~s$^{-1}$. 

In contrast to $K_2$, the primary component velocity amplitude $K_1$ 
appeared deceptively easy to determine in both binaries analyzed here.
The primary profiles seemed symmetric and 
straightforward to approximate using the classical rotational profile. 
However, it is unclear whether the  profile center accurately 
traces the motion of the  mass center, due to the distortion
of the primary component. In both binaries, 
the small $v \sin i$ width variation due to the symmetric term 
of the simplfied Roche potential expansion, $J=2$ (e.g.\  Eq.(2) in \citet{Rci1969})
was estimated at $\le\!5$ km~s$^{-1}$  (item \#2 in the list in Section~\ref{both}).
The first asymmetric term ($J=3$)  is expected to be smaller by a factor of
roughly $r/a$ compared to the dominant  tidal elongation term, %($J=2$), 
 where $r$ is the primary mean radius and $a$ is the mass center sepraration. 
The expected shift, 
%approximately one-third of the dominant distortion,
$<\!1.5-2.0$ km~s$^{-1}$,
would lead to a systematic error $\delta (q_{\rm sp}) \simeq 0.007$. Notably,
due to the sense of the Roche-lobe asymmetry, this error would be
opposite to that suggested by small value of 
$q_{\rm ph}$, i.e.\ the actual $K_1$ of the mass centre (and thus $q_{\rm sp}$)
would be  {\it larger\/} than  estimated for the symmetric profile.

 % ====== end === Section Discrepancy inf q =====================================

% =============   Section: Model  =================================   Sec.7
\bigskip

\section{The W~UMa-type binary model}                    % energy transport
\label{model}

\subsection{The Lucy model}   % ----------------------------------------------------------------   Sec.7.1
\label{mod-lucy}

When it appeared more than half a century ago, 
the \citet{Lucy1968a,Lucy1968b} model was the first consistent 
description of W~UMa binaries as dynamically stable 
configurations of solar-type stars in strong physical contact. 
The Lucy publication left important structural and energy-related 
issues open, but the light-curve calculation prescription 
was very convincing and simple: 
it suggested that common equipotentials of the Roche binary model
as better describing  components of  W~UMa binaries than -- then used --
complex descriptions of large tidal distortions imposed on spherical stars.
While the unprecedented idea of two different-mass stars 
touching each other -- and still in equilibrium -- was not accepted immediately, 
the model found enthusiastic approval among binary-star observers 
interpreting light curves. The model perfectly reproduced the 
observed light curves, assuming that both stars had identical
surface brightnesses. The particularly successful work
on totally eclipsing W~UMa-type binaries by 
\citet{MD1972a,MD1972b} followed by similar investigations 
by  \citet{Lucy1973} and \citet{WD1973} showed 
the great potential in that domain. The Lucy prescription found widespread use 
when ready-made, easy-to-install codes became widely available thanks to
the generosity of several investigators 
\citep{WVH2009,Prsa2016,2020ascl.soft04004W,Conr2020}. 
 There were voices of a rather 
low information content of the light curves \citep{Rci1993,Rci2001}
and of a sensitivity to the ubiquitous  presence of third stars \citep{Rci2007},
but they were ignored.
          
%We disregard the matter of presumed thermal cycles 
%\citep{Lucy1976,Flan1976} and -- if the cycles indeed exist --
%discuss the contact phase observed for AW~UMa and $\epsilon\,$CrA.

Interpretation of the W~UMa-type binary light curves unambiguously shows that 
the surface brightnesses of both components are almost perfectly the same,
despite commonly observed mass ratios different from unity, 
sometimes as small as $q \simeq 0.1$, as for our stars or smaller
\citep{Wadh2021,Li2022,Guo2023}.
For Main Sequence stars, the ratio of nuclear luminosities 
scales as $L_2/L_1 \propto q^\alpha$ with $\alpha \simeq 4 - 5$ so that
the secondary components in most cases would be expected to be energetically
inert stars. To explain the observed equal-temperature property,
\citet{Lucy1968a} proposed the turbulent, gravity-driven convection -- 
known for its high efficiency in solar-type stars -- 
as a process carrying colossal amounts of
energy from the primary to the secondary component. 
This assumption was entirely {\it ad hoc\/} and most likely incorrect: 
the force driving convection is directed perpendicularly to the direction 
of  energy transport in the volume between the two stars; therefore,
turbulent convection is expected to be entirely 
extinguished there or to carry insignificant amounts of energy. 
A more likely seemed to be an organized flow deep in the stellar interior
as suggested by \citet{Webb1976,Webb1977}; however, existing 
descriptions still did not include the weak, but ever-present Coriolis force. 
In both mechanisms, by Lucy and by Webbink, 
an assumption was made of energy transport originating 
in the primary component core and depositing the energy 
deep inside the secondary. 
Thus, both can be called the ``internal'' energy-transport solutions. 

%The Lucy model, by assuming the strict Main Sequence mass-radius relation
% had another weak point of requiring structurally {\it dissimilar\/} components, 
% somewhat artificially explained by \citet{Lucy1968a} 
% as caused by a difference in metallicities in Main Sequence stars. 

An ``external'' process was proposed by  \citet{Shu1976}. 
The authors pointed out that the faster evolutionary 
expansion of the more massive component may lead to a 
``boiling over'' of the primary-component matter onto the secondary,
resulting in a complete enshrouding of the secondary component 
by the hotter gas. The same matter would then be visible over the whole 
binary surface. 
This would perfectly explain the observed temperature equality, but --
contrary to normal conditions in external layers of stars -- 
the hot material would have to stably stay on the cooler gas
of the secondary.  
The inverted-temperature solution was strongly criticized as  
containing deep physical problems and has been abandoned 
\citep{Papa1979,Shu1979}.

The review by \citet{Smith84} summarized the
theoretical difficulties of the Lucy model, particularly the over-constrained
nature of its structural assumptions and the lack of a 
credible energy-transfer mechanism. 
Following remarks by \citet{Webb2003} on the lack of 
new theoretical research, subsequent investigations primarily focussed  
on binary evolution issues. \citet{YE2005} presented a comprehensive
discussion of differential evolution as a factor providing stability for
W~UMa-type systems consisting of stars of different masses. 
However, research on  energy transport between
the components has stagnated, even as the Lucy photometric 
model has continued to successfully reproduce large numbers of observed 
light curves. 
Through its frequent and extensive use, all aspects of the 
Lucy model gradually gained general acceptance, and
the unresolved issue of energy transfer ceased to be discussed.

%\bigskip

\subsection{The St\c{e}pie\'{n} model}   % --------------------------------------------------------  Sec.7.2
\label{mod-kst}

%St\c{e}pie\'{n}       -  spelling

% rewritten
The  \citet{Step2009} model of W~UMa-type binaries               
focuses on an aspect largely ignored in the Lucy model: 
the solution for the energy flow between the components with
a particular attention to the Coriolis force, which organizes the flow
and makes it concentrated to the equatorial regions. 
The model predicts inter-binary velocities so
that the Roche equipotentials cannot be defined. 
However, for sufficiently small velocities, gravity and centrifugal
forces dominate, restoring the geometrical properties 
often described by the term of the ``Roche lobes''.
% thereby undermining the main assumption
%of the Roche equipotentials, which is central to the Lucy photometric 
%model and the basis for photometric determinations of  $q_{\rm ph}$. 
%While Roche equipotentials lose their meaning, some regions
%of  gravitational dominance will still exist for moderate internal velocities.
%We are using  the term ``Roche lobes'' in this sense -- as a convenient
%construct, similar to how it has already been employed in binary evolution models --
%but not as strict, well defined surfaces. 
The St\c{e}pie\'{n} model offers a wide range 
of solutions for W~UMa-type binaries formed
from differentially evolved components and observed as various subclasses 
of such systems \citep{Step2013}.

% ------------------  Figure:  the flow model for q=0.10  ----------------------  Fig.9 
\begin{figure}[h]
\begin{center}
\includegraphics[width=11.5cm]{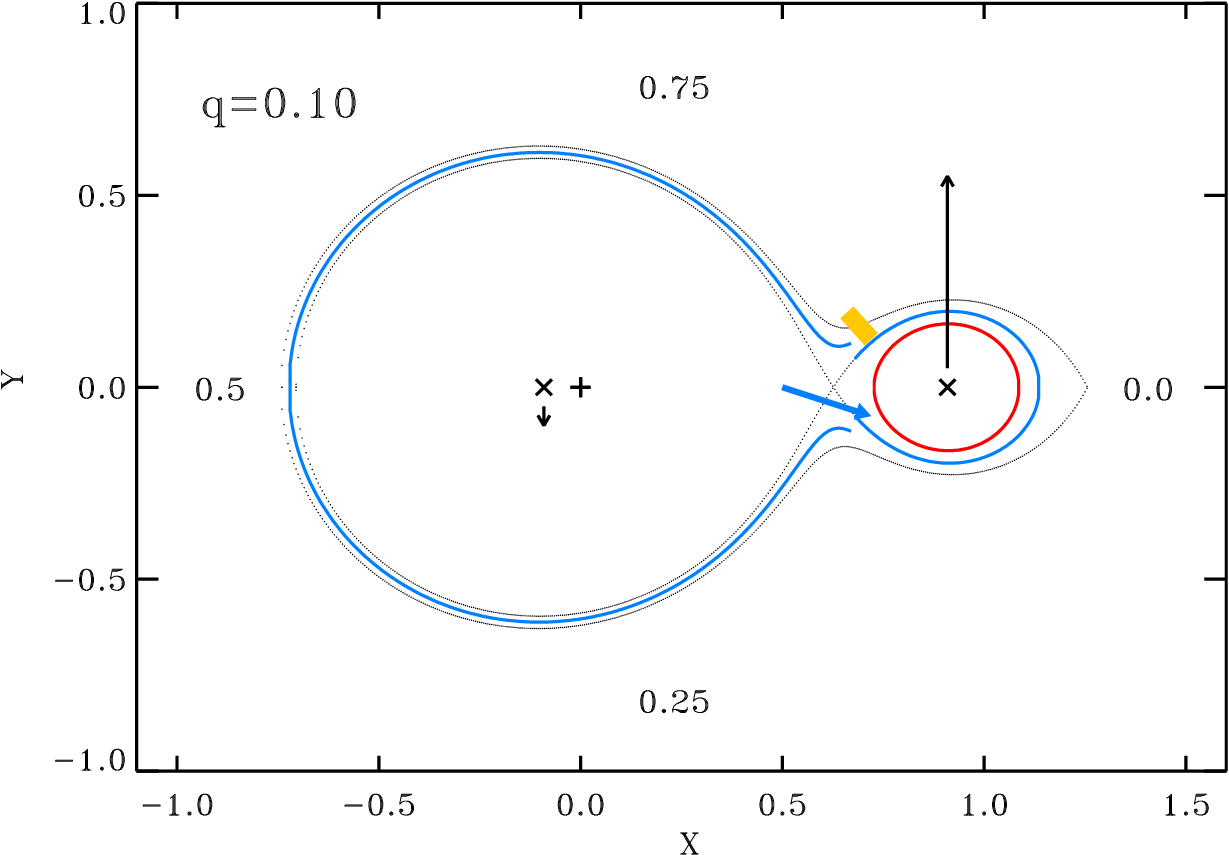}       %  equat. plane schem. flow Stepien
\caption{
\footnotesize
Properties of the \citet{Step2009}         %St\c{e}pie\'{n} 
model follow consideration of the initial stages of the Roche-lobe overflow
by the primary. The flow is deflected by the Coriolis force to the side 
of the secondary, building up a thick belt. The shape of the belt 
in the orbital plane may be close to an equipotential but different from that describing 
the primary component and very likely tighter. As a result,
one component slightly overfills and the other slightly underfills the Roche model
common equipotential. 
The figure is for $q=0.10$, similar to that of AW~UMa and $\epsilon\,$CrA. 
The blue line shows the hot gas of the expanding primary, while 
the secondary component is shown in red. The approximate location of the prominent
Enhanced Spectral-line Perturbation (ESP) observed in both binaries is marked in 
orange. 
The numbers at the perimeter give the orbital phase positions of an external observer.
The arrows at the mass centers represent (in scale) the orbital velocities of the 
component stars.
}
\label{fig_Stepien}
\end{center}
\end{figure}
% --------------------------------------------------------------------------------------------

% ------------------  Figure:    Image of Ers Cra in  KSt model   ---------------  Fig.10
\begin{figure}[h]
\begin{center}
\includegraphics[width=12.5cm]{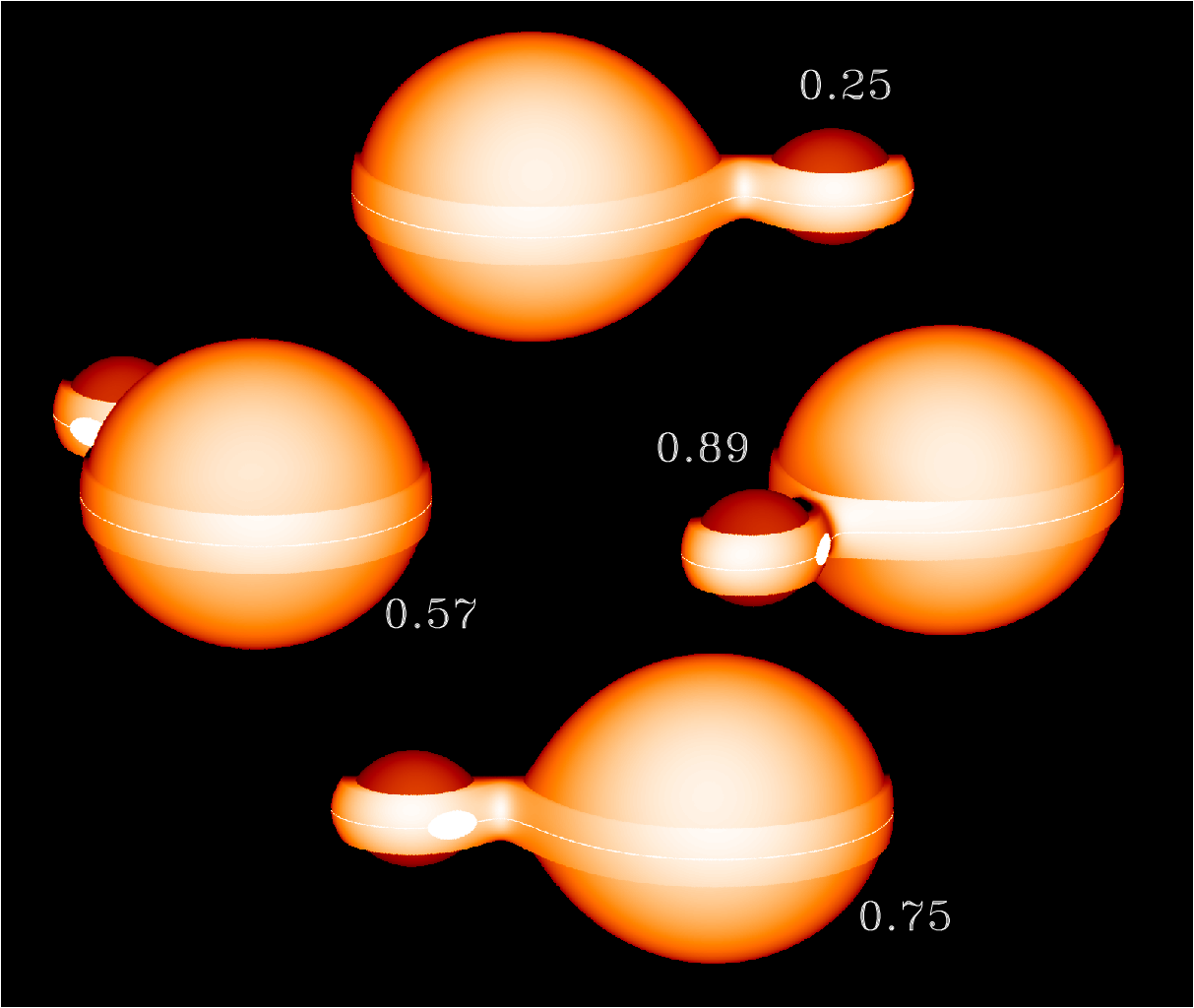}       %  image Eps CrA:  Stepien
\caption{
\footnotesize
A schematic view of $\epsilon\,$CrA at four orbital phases envisaged for
the St\c{e}pie\'{n} model; compare it  with Figure~\ref{img_EpsCrA_Lucy}
for the Lucy model. The belt surrounding both stars is pictured here with 
an arbitrary height of $\pm 0.1$ orbital separation units; its 
temperature is set slightly higher than at the side of the primary for better
visibility while the secondary is cooler.
For a simpler visual representation, all shown surfaces follow the 
equipotentials: 
the primary is at the inner critical one, the secondary at the one with the side 
radius 0.95 of its inner critical, while the belt is positioned at a half-way 
between the inner and outer critical equipotentials.  
%The secondary is shown cooler than the 8500K primary and the belt.
}
\label{img_EpsCrA_Stepien}
\end{center}
\end{figure}
% ------------------------------------------------------------------------------------

As a starting point, the \citet{Step2009} model utilizes results from
detailed hydrodynamical calculations by \citet{Oka2002}, who 
analyzed the velocity field on a Roche-lobe-overflowing component 
in a close binary system. The original calculations involved a different
scenario than a W~UMa binary:
a mass-losing component in a cataclysmic binary transferring matter
into the deep potential well of a massive, very small (collapsed) star. The 
important hydrodynamical details of the initial stages of the
outflow are expected to be very similar, irrespective of the nature of the mass
gainer. In the case of the St\c{e}pie\'{n} model, the binary components
have been brought into contact by magnetic wind and tidal
dissipation processes. The mass-losing primary is assumed to 
be a Main Sequence star, possibly showing some evolution. In fact,
the primaries of the AW~UMa and $\epsilon\,$CrA systems clearly
show a different advancement in the evolution. The binary is, therefore, of the
SD1 type in the classification of \citet{YE2005}.

The primary transfers its matter to the less-massive secondary, 
which is expected not to  expand excessively during the process. This  
suggests that the secondary is a low-mass star or the core of a mildly 
evolved MS star, possibly resulting from a previous binary mass exchange or  a
component exchange in a triple system. The stream from the primary is 
massive and forms an equatorial ``bulge'' on the secondary. Its equatorial
extension is subject to the binary gravitational forces, allowing it to fill the 
entire critical lobe of the star. 
As observed more clearly in the case of $\epsilon\,$CrA (Figure~\ref{fig_Eps_spot}),
the flow appears to exhibit a range of rotational velocities similar to that    
of a Roche-model-filling secondary at phases around 
$\phi \simeq 0.25$, but it settles to a smaller velocity range  
at the $\phi \simeq 0.75$ half of the orbit. We note that the secondary component's
signature is very difficult to observe leading to a particularly large 
measurement scatter in Figures~\ref{fig_Eps_spot} and \ref{fig_AWUMa_spot}. 

While the photospheric sound velocity for F-type stars, such as the 
AW~UMa and $\epsilon\,$CrA primaries, is of the order
7 km~s$^{-1}$, the energy-carrying flow originates in sub-photospheric layers, where 
velocities may reach 30 -- 50 km~s$^{-1}$. These velocities
remain small relative to  free-fall or orbital velocities, so  gravitational
forces  still dominate in controlling the geometric properties of the flow
and the overall geometry does not deviate appreciably from the Roche model.
The flow is confined to the equatorial regions.
It continues around the secondary and returns to 
the primary. At all phases, the matter from primary component dominates the 
external view of the binary. The actual secondary star may be almost
entirely hidden within the flow. The changing orbital aspect and eclipses of
the flow are observed as the W~UMa-type variability. The binary would be expected
to look very differently at large inclinations, but in such cases,
it would not be detected as a variable star.

Upon leaving the primary, the gas flow -- starting with moderately supersonic 
velocities -- is deflected by the Coriolis force to the side of the secondary.
This area may be visible as a hot spot and manifest itself as the O'Connell effect 
in the light curve. The binary V361~Lyr \citep{V361Lyr} 
is currently the best-known example where the deflection of the stream and
interaction with the local matter is strongly manifested. 
This side of the binary corresponds to that visible at $\phi = 0.25$ in 
AW~UMa and $\epsilon\,$CrA (Figure~\ref{fig_Stepien}). 
 The flow and its impact  appear to have been
 also observed at a moderate spectral resolution 
 in the massive W~UMa binaries strongly showing the O'Connell 
 effect, such as DU~Boo and AG~Vir, as noted by \citet{Prib2011}. 
 
%  AW~UMa and $\epsilon\,$CrA,

Most of the transferred matter completes only a single revolution
around the secondary before striking the primary's atmosphere 
on the side visible in the $\phi = 0.75$ hemisphere. 
This is where we observe the prominent Enhanced Spectral-line Perturbation (ESP), 
which is visible in both binaries discussed in this paper,
projecting against the secondary profile at the sub-observer at the same phase,
$\phi \simeq 0.65$ (marked in orange in Figure~\ref{fig_Stepien}).
It is important to stress that we see the ESPs as regions of an 
increased line-of-sight atom density, rather than as hot or cool spots. 

%  grammar checked with super model  -->

While part of the flow may end up on the primary component at that point,
the flow may continue around both components and may 
 be visible as the ``pedestal'' of the primary-component profile
(item 1.\ in the list in Section~\ref{both}). Thus, at all phases,
the light curve of a 
%low-$q$ 
binary is dominated by two sources:
the primary itself and the flow of matter encompassing the whole binary, 
originating from the primary. The secondary may be entirely 
invisible, or it may heat up to an equilibrium with the overlying equatorial belt
due to a '`bottling-up'' effect of its own energy,
constrained by a narrow polar opening, as described by \citet{Step2009}. 

As described above, the flow retains relatively 
low velocities compared to those needed to restore the dynamical 
equilibrium in the binary. Its motion around the secondary continues 
to be subject of energy and angular-momentum 
losses due to very strong tides within the secondary component's lobe.
These forces should extract the kinetic energy of the flow and
send the material ``down'', closer to the surface of the 
secondary component. Thus, the  model offers a 
prediction which may be observed as the  
 $q_{\rm ph}$ vs.\ $q_{\rm sp}$ discrepancy (Section~\ref{discr}):
While typically, we expect that the primary component  
{\it to exceed\/} the critical-lobe dimensions for the given
mass ratio, the transferred matter may end on an orbit around the 
secondary of {\it smaller dimensions\/} than the critical lobe 
(Figure~\ref{fig_Stepien}). 
This would be only a {\it tendency\/}, depending on the stage 
of binary evolution for a particular case and the intensity of the 
mass flow.
It is important to stress that, in the case of the new model, 
the photometric mass ratio are determined
from the shape of the flow, not from dimensions of
the underlying stars. Thus, any variations of 
$q_{\rm ph}$,  as described by \citet{ND1991} for AE~Phe,
would probably indicate variations in the shape or mass-loading
of the flow. 

The \citet{Step2009} model may find particularly fertile ground for 
spectroscopic high-resolution investigations among the 
``poor thermal contact binaries'', i.e.\ , binaries as close as those of the
W~UMa-type but with components of unequal temperatures.
In such binaries, the flow is clearly unable to transfer
the energy between the components. These binaries are usually the most
unstable, as if seeking a new equilibrium state. 
\citet{Siw2010} conducted a combined photometric/spectroscopic 
investigation of several of such binaries using the best currently available 
techniques.  
The spectroscopic data show well-defined profiles of cool and faint 
secondary components and the presence of strong ESP features visible
on the $\phi = 0.75$ hemisphere. CX~Vir was most
extensively observed: The  ``phase 0.65'' ESP feature in that binary 
does not seem to look the same as in AW~UMa and $\epsilon\,$CrA; it is much 
stronger and seems to show a different phase-drift dependence. 
Firm conclusions from the excellent CX~Vir data are hard to derive
because the sparse  phase coverage in the \citet{Siw2010} paper 
is limited to narrow intervals around
the orbital quadratures. We can recognize the features interpretable
by the St\c{e}pie\'{n} model, but the details seem to be different. 
 
 % ----------------------------end  KSt model -------------------------------------

% Hydrodynamic simulations similar to those of \citet{Oka2002}  are very much
% needed for the model of \citet{Step2009}.  

 % === end Section:   Model ================================================

% Section ======================   Conclusions =====================  Sec.8

%\bigskip
%\vfill

\section{Conclusions}
\label{concl}

The paper contains a combined discussion of the previously obtained results for the
two W~UMa-type binaries, AW~UMa (P1) and $\epsilon\,$CrA (P2).
Both binaries were monitored over time at high spectral resolution, 
and they share many physical properties, notably the very small 
mass ratio and early-F spectral type. Their similarity largely stems from the
availability of instrumentation but has revealed interesting and
unexpected details discussed in this paper. 
Additionally, the later analysis of $\epsilon\,$CrA, which is a simpler binary, 
helped to interpret unexplained features observed in AW~UMa in P1. 

Both binaries -- particularly AW~UMa -- have been repeatedly observed photometrically,
and their light curves perfectly agree with the common-equipotential 
light-curve-synthesis model of \citet{Lucy1968a,Lucy1968b}. 
However, the spectroscopic perspective presents a more complex picture of
%the binaries are not rotating as solid bodies, as required by the strict
%Lucy model, but instead exhibit several phenomena related to 
a circumbinary flow with velocities not exceeding 30 -- 50 km~s$^{-1}$.
Such velocities, moderate compared to the free-fall or 
orbital velocities of about 300 -- 350 km~s$^{-1}$, 
are predicted in the \citet{Step2009} energy transport model of W~UMa binaries.
This model envisages 
%the energy transport between the binary components as 
an optically thick 
stream or belt of matter originating from the Roche-lobe-overfilling 
primary component, which retains its heat during the transport (Section~\ref{mod-kst}). 
The observed W~UMa-type variations result from the changing visibility 
and eclipses of this energy-carrying belt, whose eclipse and visibility-aspect 
 variations  closely resemble those of the Roche model geometry.

%The velocity field around the secondary component carries important information 
%relevant to the \citet{Step2009} model. 
We note (Figure~\ref{fig_Eps_spot}) that the observed
rotational velocities of the $\epsilon\,$CrA secondary 
are very similar to those expected for a rigidly 
rotating Roche model in the first half of the
orbit ($\phi = 0.25$). In contrast, the second half of the orbit ($\phi = 0.75$) 
shows possible
 settling of the velocities to lower rotational velocities. This behavior 
is exactly as predicted for a flow that initailly fills the critical lobe and then 
 loses its energy, drifting closer to the secondary's otherwise
 invisible  surface. The same effect is not as clearly visible in the much more
complicated case of AW~UMa (Figure~\ref{fig_AWUMa_spot}), which was also
a less-well observed binary. 

The new results for AW~UMa confirm  
the ``large'' value of the spectroscopically determined mass-ratio 
$q_{\rm sp} = 0.092 \pm 0.007$ (Section~\ref{sec-aw}),
in contrast to the often quoted photometrically derived value, 
$q_{\rm ph} = 0.080 \pm 0.005$ 
(this error comes from a scatter of several independent solutions).
The new determination agrees with the previous spectroscopic result obtained
at a medium spectral resolution, $q_{\rm sp} = 0.101 \pm 0.006$ 
\citep{PR2008}. The difference in spectral resolutions is important, as it
relates to the different ways of measuring the secondary component's 
orbital semi-amplitude $K_2$. The high spectral resolution
result is based on the very weak, almost flat profiles of the AW~UMa secondary,
with an attempt not to include 
the Enhanced Spectral-line Perturbation (ESP) features. In contrast,
the previous medium-resolution measurements included the then-unrecognized 
ESPs in the averaged secondary profiles. The ESPs are not hot or cool
regions but rather volumes of increased line-of-sight densities of atoms,
having excitation properties similar to those of the primary component's outer regions.
The AW~UMa secondary typically shows two to four such localized ESPs, 
which are seen migrating as the binary rotates, projecting onto the 
secondary profile.
In both binaries, one strong ESP was localized in the region between the stars,
projecting on the secondary component profile at the sub-observer
orbital phase $\phi = 0.65$. Such location is expected in the St\c{e}pie\'{n}
model as a place where the returning stream collides with the primary component
after revolving around the secondary. 

% two first sentences rewritten 
The difference between $\epsilon\,$CrA and AW~UMa in the number of 
ESPs currently lacks an explanation. While the ESPs are most likely driven by local 
thermodynamical instabilities and are chaotic in nature, it is possible
that the two binaries differ in the type or extent of the flow.
The stronger ESP variability in AW~UMa may indicate 
that the flow around its secondary, as envisaged in the St\c{e}pie\'{n}
model, is more erratic than that in $\epsilon\,$CrA or is
subject to a localized pileup. We also note  opposite tendencies in 
the systematic orbital-period changes of the two binaries (Section~\ref{both}):

% rewritten slightly for clarity
Based on the sign of the period changes, 
the {\it net mass exchange\/} between the components of AW~UMa appears 
to be directed from the more massive primary to the less massive secondary, 
whereas in $\epsilon\,$CrA, it flows in the opposite direction. 
One possible explanation is that the multitude of ESPs on the AW UMa 
secondary signals an "overload" of the flow, caused by the abundance of
incoming matter at this particular stage of binary evolution

The confirmation of the  $q_{\rm ph} < q_{\rm sp}$ inequality in AW~UMa, along
with the absence (or small size) of a similar discrepancy in  $\epsilon\,$CrA,
prompted an analysis of several binaries with the best-constrained
photometric solutions in Section~\ref{discr}. A careful selection of W~UMa 
binaries with spectroscopic $q_{\rm sp}$ determinations and the best (total eclipse)
Lucy-model photometric solutions indicates that discrepant results 
are common and that smaller values of  $q_{\rm ph}$ are a persistent tendency 
(Section~\ref{discr}, Figure~\ref{fig_q-discrep}). 
This could be explained within the \citet{Step2009} model 
as due to the tight circulation of the energy-transporting belt around
the secondary component, in contrast to the Roche-lobe 
overfilling primary. Thus, the two stars are expected to show opposite 
tendencies for deviations from the common equipotential of the Lucy model;
this, in turn, could lead to a smaller value of $q_{\rm ph}$ than the real one.

%By explaining the energy-transfer problem
%in  W~UMa-type binaries and providing possible reasons 
%for the $q_{\rm ph} < q_{\rm sp}$ tendency, the St\c{e}pie\'{n} model 
%casts doubt on the applicability
%of the common-equipotential Lucy model for  deriving physical
%data for these binaries. 

While the St\c{e}pie\'{n} model may suggest directions for exploring 
the $q_{\rm ph} < q_{\rm sp}$ tendency, it 
%Yet, the new model 
does not provide  predictions regarding the magnitude of deviations from
the Lucy model or  suggest methods to reconcile 
$q_{\rm ph}$ or $q_{\rm sp}$ results. 
%Are the results already obtained with the photometric Lucy model still useful? 
Currently, the photometric approach dominates,
%This question applies to large-scale projects 
utilizing complex, state-of-the-art light-curve synthesis codes
\citep{Prsa2016,Conr2020,2020ascl.soft04004W}        % ,Lat2021}
%for the currently ongoing large photometric surveys, following the example
which have largely superseded earlier explorations 
based on based on ASAS \citep{Pil2009},   % Pilecki (PhD 2009)
OGLE \citep{Sosz2015}    %  Soszynski et al. 2015
surveys, and the extensive compilation of nearly 700  results by 
\citet{Lat2021}.
A convenient tool for a large-scale handling of statistical $q_{\rm ph}$
data -- implicitly assuming a strict acceptance of the Roche model --
has already been developed by \citet{Pejcha2023}.
%Another question is: 
But do summaries combining $q_{\rm ph}$ and 
$q_{\rm sp}$ data, such as those by \citet{GS2008}, \citet{Lat2021} 
or \citet{Gaz2024}, provide an unbiased picture?

%While the Roche-model ``short cut'' to massive determination of $q$ 
%appears to posses a serious difficulty in providing biased results,
%we currently have no reasons to expect any similar 
%bias in spectroscopic results. 
New spectroscopic investigations of W~UMa-type binaries are urgently needed. 
Observations of AW~UMa at two spectral resolutions 
suggests that  medium-resolution 
($R \simeq 15,\!000 - 20,\!000$) data may suffice for an acceptable 
determination of $q_{\rm sp}$. However, detailed analysis of
spectral profiles and detection of ESPs require 
a higher resolution ($R > 30,\!000$).  The binary FG~Hya 
($V \simeq 10$, $q_{\rm sp} = 0.14$) 
is the only case of the inverted mass-ratio discrepancy  
(Section~\ref{discr}, Figure~\ref{fig_q-discrep}), and it urgently requires a new
spectroscopic determination. Binaries
with less extreme mass ratios particularly need spectroscopic work:
we already have tantalizing indications \citep{DDO-11,PR2008} 
that medium-resolution RV profiles for V566~Oph 
($V \simeq 7.5$, $q_{\rm sp} = 0.263 \pm 0.012$)     % ; Table~\ref{tab_Qtab})  
look -- unlike those of our two binaries                      %  AW~UMa or $\epsilon\,$CrA 
-- as predicted by the Lucy model.
Interpreted within the St\c{e}pie\'{n} model, this may indicate 
that the circumbinary flow fully dominates the spectroscopic
profile shape for V566~Oph, in contrast to our two,
very low mass-ratio systems where the primaries are dominant. 
It is possible that the flow dominance is a typical situation for W~UMa binaries
with mass ratios larger than some specific value. We note 
that the two binaries considered here are not  cases of  extremely 
small $q_{\rm ph}$, as smaller values have  already been reported in the 
literature \citep{Wadh2021,Li2022,Guo2023}. 

At the other end of the mass ratio range, 
two moderately bright W~UMa binaries would be
particularly important for a broader spectroscopic picture:
SW~Lac ($V \simeq 8.7$) with $q_{\rm sp} = 0.776 \pm 0.012$ \citep{DDO-10} 
and OO~Aql ($V \simeq 9.5$) with $q_{\rm sp} = 0.846 \pm 0.007$ \citep{DDO-12}.
We noticed during the DDO program that the RV profiles of SW~Lac 
have shapes very much like those of a genuine semi-detached system (SD2), 
yet with components having identical atmospheric properties. 
Finally, the poor thermal contact 
binaries (Section~\ref{mod-kst}), particularly the bright ones 
previously analyzed by \citet{Siw2010}, may --
when observed spectroscopically over all orbital phases --
show more details of the flow. Possibly the flow is too weak 
to engulf the equatorial regions of 
the secondary yet strong enough to produce the already observed,
prominent ESP features. 

For a final summary:
%The simplistic Roche-model light-curve calculations have not significantly 
%advanced our understanding of W~UMa binaries. 
While spectroscopy is essential for our understanding of W~UMa binaries,
it remains difficult and costly. 
We have some insight into how these binaries form in small 
stellar groups and that they likely consist of differentially evolved components.
However, we are still in the early stages of understanding the structure 
and surprising stability of these binaries. 
The roles of tidal energy and angular momentum dissipation 
remains to be studied and incorporated into models similar 
to those of \citet{Oka2002}. 
Such models could help in interpreting spectroscopic results, partly regarding 
the complex velocity fields observed within these binaries.

% St\c{e}pie\'{n} 
% \citet{Step2009} model to the flows observed in AW~UMa and $\epsilon\,$CrA.

 % ======== Ending ===========================================================

%\bigskip

\begin{acknowledgements}

I would like to thank Dr.\ Andrej Pr\v{s}a who, acting as a reviewer of the paper, suggested
 importnat improvements and corrections to the text.

I would like to thank Staszek Zo{\l}a and Micha{\l} Siwak for useful suggestions 
and comments.

\medskip
Special thanks go to Kazik St\c{e}pie\'{n}, the originator of the new model for W~UMa
binaries for his insightful ideas, interesting discussions, and suggestions on the presentation
of  arguments. 

\end{acknowledgements}

\newpage
%\vfill
%\bigskip

% ========      The triple-system formation hypothesis for AW UMa and Eps CrA 
% not discussed at all

% ===========================================================================

% \newpage

\end{document}